\documentclass{aa}  
\usepackage{graphicx}
\usepackage{amsmath}
\usepackage[]{natbib}
\bibpunct{(}{)}{;}{a}{}{,} 
\usepackage[varg]{txfonts}
\begin{document}

   \title{Multi-zone non-thermal radiative model for stellar bowshocks} 

   \titlerunning{Non-thermal radiation from stellar bowshocks}
   \authorrunning{S. del Palacio et al.} 


   \author{
          S. del Palacio \inst{1,2} 
          \and V. Bosch-Ramon \inst{3}
          \and A. L. M\"uller\inst{1,4,5} 
          \and G. E. Romero\inst{1,2} 
          }
          

   \institute{
           Instituto Argentino de Radioastronom\'{\i}a (CCT La Plata, CONICET), C.C.5, (1894) Villa Elisa, 
            Buenos Aires, Argentina.\\ \email{[sdelpalacio,almuller,romero]@iar-conicet.gov.ar}
          \and Facultad de Ciencias Astron\'omicas y Geof\'{\i}sicas, Universidad Nacional de La Plata, Paseo del Bosque, B1900FWA 
            La Plata, Argentina.
          \and Departament de F\'isica Qu\`antica i Astrof\'isica, Institut de Ci\`encies del Cosmos (ICCUB), Universitat de Barcelona (IEEC-UB), Mart\'{\i} i Franqu\`es 1, E-08028 Barcelona, Spain.\\ \email{vbosch@am.ub.es}
          \and
            Institut f\"ur Kernphysik (IKP), Karlsruhe Institute of Technology (KIT), Germany. 
          \and 
            Instituto de Tecnolog\'{\i}as en Detecci\'on y Astropart\'{\i}culas (CNEA, CONICET, UNSAM), Buenos Aires, Argentina.
             }

   \date{Received ; accepted }

 
  \abstract
   {Runaway stars produce bowshocks that are usually observed at infrared (IR) wavelengths. Non-thermal radio emission has been detected so far only from the 
   bowshock of \object{BD$+43^{\circ}3654$}, whereas the detection of non-thermal radiation from these bowshocks at high energies remains elusive.}
   {We aim at characterising in detail the radio, X-ray, and $\gamma$-ray emission from stellar bowshocks accounting for the structure of the region of interaction between the stellar wind and its environment.}
   {We develop a broadband-radiative, multi-zone model for stellar bowshocks that takes into account the spatial structure of the emitting 
   region and the observational constraints. The model predicts the evolution and the emission of the relativistic particles accelerated and
   streaming together with the shocked flow.}
   {We present broadband non-thermal spectral energy distributions for different scenarios, synthetic radio-cm synchrotron maps that 
   reproduce the morphology of \object{BD$+43^{\circ}3654$}, and updated predictions in X-ray and $\gamma$-ray energy ranges. We also compare the results of the multi-zone model applied in this work with those of a refined one-zone model.} 
   {A multi-zone model provides better constraints than a one-zone model on the relevant parameters, namely the magnetic field intensity and 
   the amount of energy deposited in non-thermal particles. However, one-zone models can be improved by carefully characterising the 
   intensity of the IR dust photon field and the escape rate of the plasma from the shocked region. Finally, comparing observed radio maps 
   with those obtained from a multi-zone model enables constraints to be obtained on the direction of stellar motion with respect to the observer.} 

   
   \maketitle
%
%


\section{Introduction}\label{sec:intro}

Massive stars have supersonic strong winds that sweep-up and heat the gas and dust from the surrounding interstellar medium (ISM), 
generating cavities known as wind-blown stellar bubbles. When these stars have a high (supersonic) peculiar velocity with respect to the ISM, the geometry of the bubble becomes bow-shaped instead of spherical \citep{Weaver1977}. The heated ambient dust and gas emit mostly infrared (IR) radiation, which can be detected thus allowing the identification and characterisation of these bowshocks (BSs) \citep[][and references therein]{NoriegaCrespo1997}

Strong BSs are promising places for the acceleration of relativistic particles, which produce peculiar radiative features 
throughout the whole electromagnetic spectrum, especially at radio-cm frequencies and at energies above 1~keV 
\citep[i.e. X-rays and $\gamma$-rays;][]{delValle2012}. However, despite the large number of such objects observed to date 
\citep{Peri2012, Peri2015, Kobulnicky2016}, \object{BD$+43^{\circ}3654$} remains the only one in which non-thermal (NT) radio 
emission has been observed \citep{Benaglia2010}. Until now, no stellar BS has been detected either
at X-rays \citep{Toala2016, Toala2017, DeBecker2017}, high-energy $\gamma$-rays\footnote{During the review of this article, \cite{Sanchez2018} published a work with a possible association of two \textit{Fermi} sources with stellar BSs.} \citep{Schulz2014}, or very high-energy $\gamma$-rays \citep{HESSColl2017}.

The radio emission at low frequencies is expected to be dominated by synchrotron radiation produced by the interaction of 
relativistic electrons with the local magnetic field \citep{Ginzburg1965}.
The detection of NT radio emission is ubiquitous in systems involving shocks, such as colliding-wind massive binaries 
\citep[e.g.][]{Eichler1993}, supernova remnants \citep[e.g.][]{Torres2003}, and proto-stellar jets 
\citep[e.g.][]{Marti1993,RodriguezK2017}. 
This suggests that shocks produced by strong stellar winds (SWs) are suitable for the acceleration of particles 
(electrons, protons, and heavier nuclei) up to relativistic energies. 
In this context, diffusive shock acceleration (DSA) \citep{Axford1977,Krymskii1977,Bell1978,Blandford1978} is the most likely mechanism. 
Furthermore, it is expected that these relativistic particles 
radiate their energy at energies from radio to $\gamma$-rays. A few models have been developed to address the NT 
emission from stellar BSs \citep{delValle2012, delValle2014, Pereira2016}, some of which over-predicted the high-energy radiation 
from these systems. 

In this work, we revisit the assumptions of previous emission models for stellar BSs and apply a 
new multi-zone emission model, presented in Sect.~\ref{sec:model}, to BSs. In Sect.~\ref{sec:results}, we present the results of applying 
this model and discuss them in order to assess future radio, X-ray, and $\gamma$-ray surveys that search for NT emission from stellar BSs. 
We also show synthetic radio-cm synchrotron emission maps that reproduce the available data and radio morphology of 
\object{BD$+43^{\circ}3654$}, and make updated predictions for this object in X-rays and $\gamma$-rays consistent with the latest observational
constraints. The conclusions are summarised in Sect~\ref{sec:conclusions}.


\section{Model}\label{sec:model}


Most of the NT radiation models presented in the literature rely on the one-zone approximation, 
which assumes that the emitter can be considered as point-like, that is, homogeneous and of irrelevant size. 
However, the validity of such models has been questioned in view of the discrepancies between predictions and observations
\citep{Toala2016}. Additionally, the structure of the BS can, in principle, be resolved with current radio interferometers 
and X-ray satellites, 
but it is not possible to address correctly issues of the spatial distribution of the emission using one-zone models.

A complete broadband radiative model of the BS needs to take into account the magnetohydrodynamics (MHD) of the stellar wind shock 
and its evolution, the acceleration of relativistic particles, and the emission of these particles considering inhomogeneous 
conditions throughout the BS. Here, we develop an extended model for the BS emitter in which we consider analytical prescriptions for the MHD, a DSA mechanism for the acceleration of particles, and a detailed numerical method for the calculation of the particle emission throughout the BS. 
The details of the model are specified below.


\subsection{Geometry}

In the reference frame of a moving star, the stellar BS is the result of the interaction of the ISM 
material acting as a planar wind and the stellar spherical wind. The shape and dynamics of stellar BSs have 
been studied by several authors \citep[e.g.][]{Dyson1975,Wilkin1996,Meyer2016,Christie2016}. 
The collision of the SW and the ISM forms an interaction region consisting of a forward shock that propagates through 
the ISM, a contact discontinuity (CD) between the two media, and a reverse 
shock (RS) that propagates through the unshocked SW. The CD is the surface where the flux of mass is zero. The reverse shock is adiabatic and fast, whereas the forward shock is 
radiative and slow \citep[e.g.][]{VanBuren1993}. Since DSA works in the presence of strong adiabatic shock waves, we will 
focus exclusively on the reverse shock. The stagnation point is located at a distance $R_0$ from the star, on the symmetry axis of the BS (with the direction of the stellar motion), and is the point where the ram pressure of the SW and the ISM completely cancel each other out. 

For a star with a mass-loss rate $\dot{M}_\mathrm{w}$, wind velocity $v_\mathrm{w}$, and a spatial velocity 
$V_\star$, moving in a medium of density $\rho_\mathrm{ISM}$, the stagnation point is at
\begin{equation}
R_0 = \sqrt{\dot{M}_\mathrm{w} v_\mathrm{w} /(4 \pi \rho_\mathrm{ISM} {V_\star}^2)}.
\end{equation}
As $R_0 \gg R_\star$, 
it is valid to simply adopt $v_\mathrm{w} = v_\infty$. Any position on the CD can be determined by an angle $\theta$ from the BS symmetry axis, which was characterised by \citet{Wilkin1996} for the case of a cold ISM (i.e. with negligible thermal pressure): 
\mbox{$R(\theta) = R_0 \csc{\theta}\sqrt{3\,(1-\theta \cot{\theta})}$}. 
\citet{Christie2016} generalised this solution for the case of non-negligible thermal pressure in the ambient 
fluid, opening the possibility to study the BSs in warm or even hot fluids such as accretion disks. 
According to their results, the width of the shocked stellar wind region at $\theta$ is well approximated as 
$H(\theta) \approx 0.2\,R(\theta)$.

\begin{figure}
    \centering
    \resizebox{\hsize}{!}{\includegraphics[width=0.5\textwidth, angle=0]{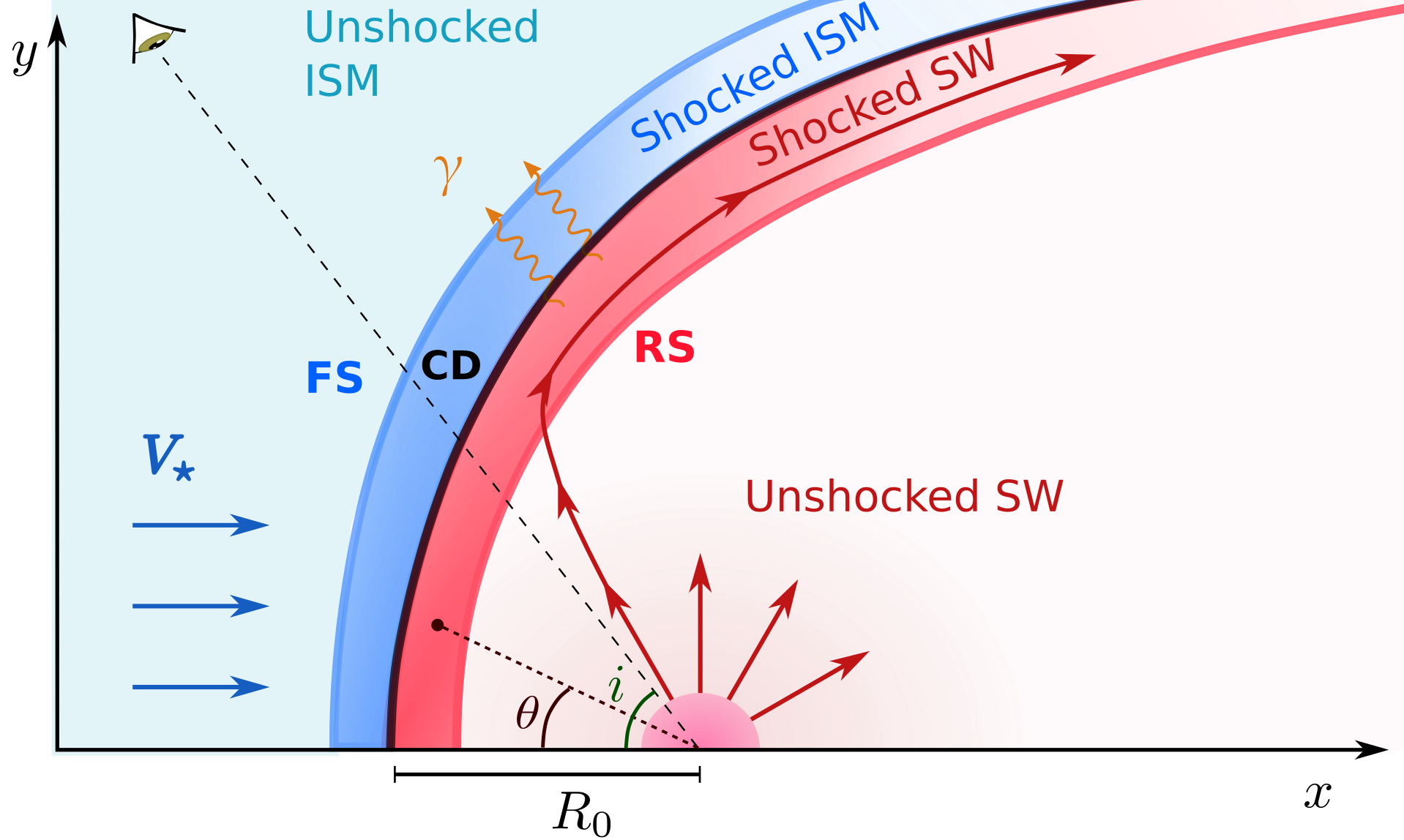}}
    \caption[]{Sketch (not to scale) of the model considered in this work. Despite the axis representation, the $(0,0)$ position corresponds to the star location. In 
    the star reference frame, the ISM moves with velocity equal to $V_\star$. 
    The position of the contact discontinuity (CD), forward shock (FS), and reverse shock (RS) are shown in solid lines. 
    The solid line with arrows coming from the star and entering into the shocked region represents one of the streamlines 
    that conform the emitter. The inclination angle $i$ and the position angle $\theta$ are also shown.}
    \label{fig:model}
  \end{figure}

Following \citet{delPalacio2016}, we develop a two-dimensional (2D) model assuming that the BS is an axisymmetric shell of negligible width. Given that the shocked gas flows at a fixed angle $\phi$ around the symmetry axis, we can restrict
most of our analysis to a one-dimensional (1D) description of a fluid moving in the $XY$ plane. A schematic picture of
the model is shown in Fig.~\ref{fig:model}. The position of a fluid element on the $XY$ plane is solely
determined by $\theta$. As the magnetic field is not dynamically relevant, it is considered that a fluid element, upon entering the RS, moves downstream from the BS apex. We assume that NT particles are accelerated once the fluid line enters the RS region, and that these particles flow together with the
shocked fluid, which convects the ambient magnetic field.\footnote{We note that the flow is not likely to be
completely laminar nor the magnetic field completely ordered, but for simplicity we neglect
in this description any macroscopic effect of the 
turbulent component of the flow and the irregular component of the magnetic field.}
  
The BS radiation is formed by a sum of these 1D emitters (linear emitters hereafter) that
are symmetrically distributed around the direction of motion of the star in a three-dimensional (3D) space: 
each discrete emission cell is first defined in the $XY$ plane, and the full 3D structure of the wind interaction zone is obtained 
via rotation around the $X$ axis, in the $\phi$ direction. 
The hydrodynamics (HD) and particle distribution have azimuthal symmetry. The dependence with the azimuthal
angle arises only for processes that depend also on the line of sight (Sect.~\ref{sec:maps}). 

We assume that the flow along the BS is laminar, neglecting mixing of the fluid streamlines or mixing between shocked
wind and medium. 
The particles are followed up to an angle $\theta\sim 135\degr$, or equivalently, until they travel a distance $\sim 5\,R_0$. 
According to our simulations, more than $50\%$ of the emission is produced within $\theta < 60\degr$ ($\sim R_0$), 
close to $90\%$ within $\theta < 120\degr$ ($\lesssim 4\,R_0$), and above $99\%$ within $\theta < 135\degr$. 
Therefore, $\theta\sim 135\degr$ is sufficient to capture most of the emission from the injected particles, and it 
also allows us to capture most of the wind kinetic luminosity available for NT particles.


\subsection{Hydrodynamics} \label{subsec:hydrodynamics}

The HD of the BS, in particular the shocked SW, depends on the star mass-loss rate, $\dot{M}$, and the wind 
terminal velocity, $v_\infty$. A stationary approximation is valid as long as $V_\star \ll v_\infty$, as the
shocked stellar wind leaves the BS before the environment can change significantly. In addition, the SW shock is adiabatic, that is, 
the shocked stellar wind leaves the BS before radiating its energy, which enhances its stability. 
Such wind shocks are expected to be quite stable for supersonic (non-relativistic) stars \citep{Dgani1996}.

We apply the analytical HD prescriptions given by \citet{Christie2016} to characterise the values of the relevant thermodynamical quantities in the shocked SW (which we assume to be co-spatial with the CD), and to obtain tangent and perpendicular vectors to the BS surface at each position of the CD. The thermodynamical quantities in the shocked SW rely on the assumption that the fluid behaves like an ideal gas with adiabatic coefficient $\gamma_\mathrm{ad} = 5/3$, and the application of the Rankine-Hugoniot jump conditions for strong shocks. The magnetic field is obtained by assuming that, at each position, its pressure is a fraction $\zeta_B$ of the thermal pressure:
 \begin{equation} 
  B(\theta) = \left[\zeta_B \, 8\pi P(\theta) \right]^{1/2}, \quad P(\theta)=\frac{2}{1+\gamma_\mathrm{ad}} \, \rho_\mathrm{w}(R(\theta))\,v_{\mathrm{w},\perp}^{2}(\theta).
 \label{eq:B}   
 \end{equation}
 
\begin{figure}
    \resizebox{\hsize}{!}{\includegraphics[width=0.35\textwidth, angle=270]{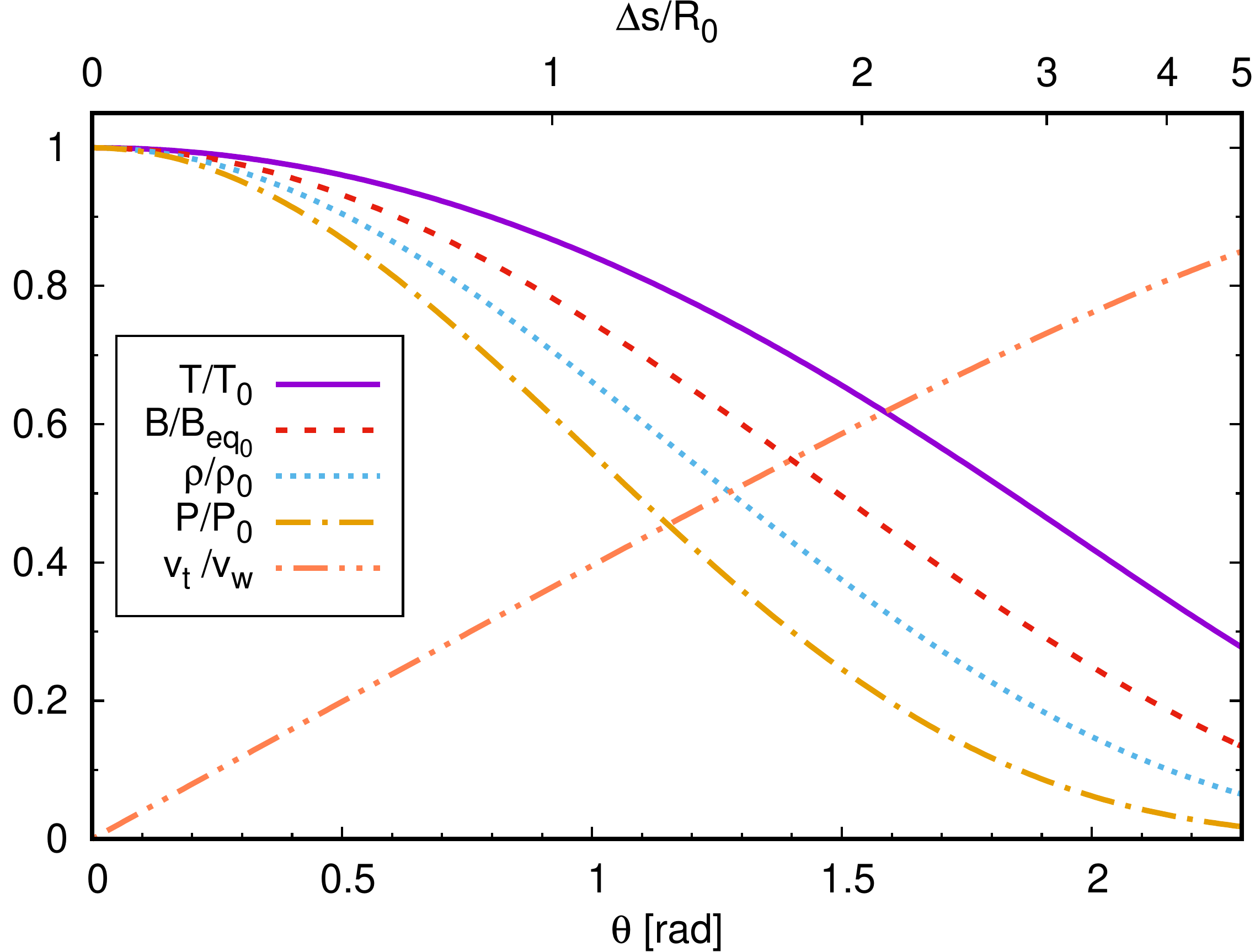}}     
    \caption[]{Thermodynamic quantities of the shocked SW: temperature (solid), magnetic field (long-dashed), density (short-dashed), pressure (dot-dashed), and tangential velocity (double dot-dashed). Quantities with sub-index zero are evaluated at the apex position (see text). The distance to the apex along the shock is given by \mbox{$\Delta s(\theta) = \int_ 0^\theta \mathrm{d} s(\theta')$}.}
    \label{fig:termo_sw}
  \end{figure}

 The dependence of the thermodynamic quantities with $\theta$ is shown in Fig.~\ref{fig:termo_sw}. Near the apex, the incoming SW 
 impacts on the BS surface perpendicularly, leading to a big jump in the gas pressure and temperature, and the fluid practically halts 
 (i.e. quasi-stagnates). As $\theta$ increases, the shock becomes more oblique, the temperature and pressure are lower, and the 
 tangential velocity becomes a substantial fraction of $v_\infty$ (in fact, the fluid can accelerate a bit further because of the 
 pressure gradient).

At large distances from the star, the stellar magnetic field is expected to be toroidal and its intensity to
drop with the inverse of the distance to the star \citep{Weber1967}. An upper limit on the stellar surface
magnetic field can therefore be estimated by assuming that the magnetic field in the BS comes solely from 
the adiabatic compression of the stellar magnetic field lines. Adopting an Alfv\'en radius 
$r_\mathrm{A} \sim R_\star$, we get \mbox{$B_\star = 0.25\, B(\theta)\, (R(\theta)/R_\star)\,(v_\infty/v_\mathrm{rot})$} \citep[][and references therein]{Eichler1993}. We note that it is possible that the magnetic
fields are strongly amplified or even generated \textit{in situ} 
\citep[e.g.][and references therein]{Schure2012}, and therefore the upper limits we derive 
for $B_\star$ could be underestimated.


\subsection{Non-thermal particles} \label{sec:distribution}
  
The BS produced by the SW consists of hypersonic, non-relativistic, adiabatic shocks, where NT particles of energy $E$ 
and charge $q$ are likely accelerated via DSA. Assuming diffusion in the Bohm regime, the acceleration
timescale can be written as 
\mbox{$t_\mathrm{acc} \approx \eta_\mathrm{acc} E \left( B \, c \, q \right)^{-1}$~s}, with 
$\eta_\mathrm{acc}\gg 1$ being the acceleration efficiency \citep{Drury1983}.
The energy distribution of the accelerated particles at the injection position is taken as 
\mbox{$Q(E)\propto E^{-p} \exp{(-E/E_\mathrm{cut})}$}, where $p$ is the spectral index of the particle energy
distribution and $E_\mathrm{cut}$ is the cut-off energy, obtained by equating $t_\mathrm{acc}$ with the minimum between
the cooling and escape timescales. The canonical spectral index for DSA in a strong shock is $p=2$.

\begin{figure}
    \resizebox{\hsize}{!}{\includegraphics[width=0.35\textwidth, angle=270]{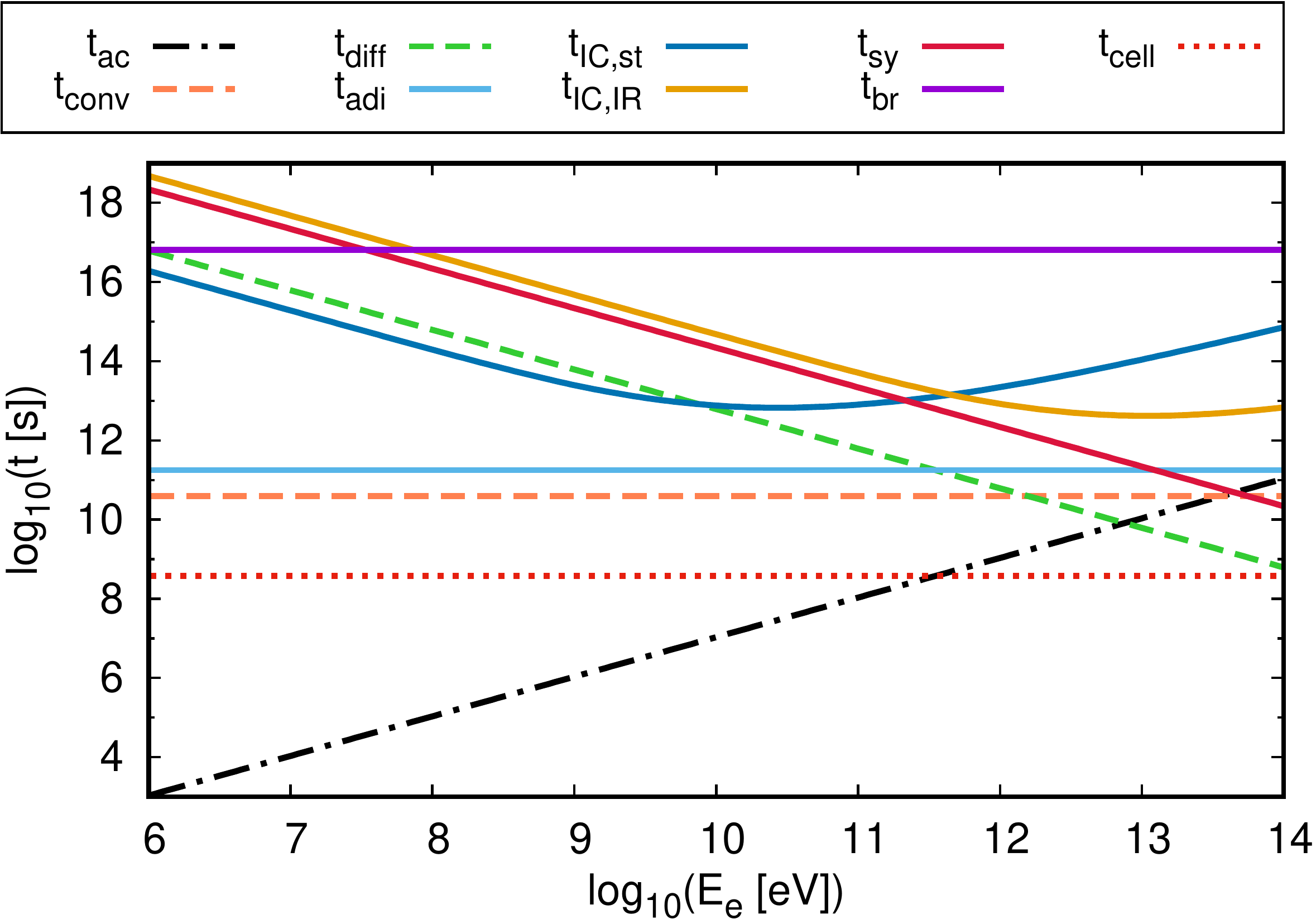}}   
    \resizebox{\hsize}{!}{\includegraphics[width=0.35\textwidth, angle=270]{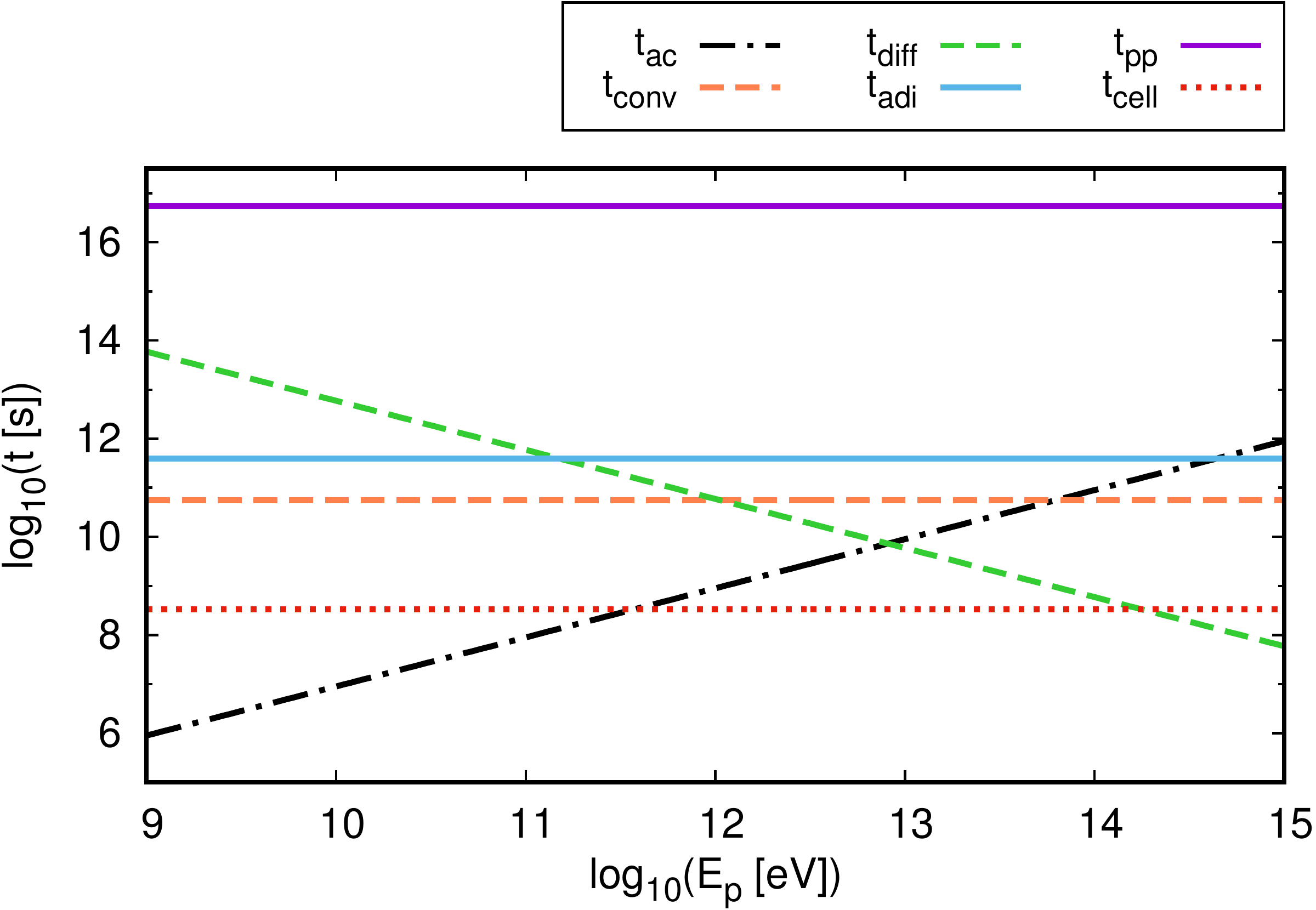}}     
    \caption[]{Characteristic cooling and acceleration times for electrons (top) and protons (bottom) at a position
    $\theta = 30\degr$ for the generic scenario presented in Table~\ref{table:parameters}. Solid lines are used for
    cooling processes whereas long dashed lines are used for escape processes. The dotted red line represents the cell
    convection time (see text in Sect.~\ref{sec:numerical}).}    
    \label{fig:tiempos}
  \end{figure}

Electrons at each cell of the BS cool through various processes, namely adiabatic losses (work done
expanding with the thermal fluid), Bremsstrahlung, synchrotron, and inverse Compton interactions (IC) with the ambient
radiation fields, namely IR from dust emission and ultraviolet (UV) photons from the star. An example of 
the relevant timescales is shown in Fig.~\ref{fig:tiempos} for the scenario discussed in Sect.~\ref{sec:one_vs_multi}.
Adiabatic losses are not dominant with respect to escape losses in this scenario because of the relatively smooth 
density evolution (see Fig.~\ref{fig:termo_sw}), although they are the dominant cooling process for electrons 
with $E_\mathrm{e} \lesssim 1$~TeV. For near-equipartition magnetic field values ($\zeta_B \sim 1$), 
radiative losses are relevant for electrons with $E_\mathrm{e} \gtrsim 100$~GeV, as synchrotron dominates their 
cooling. Otherwise, for modest magnetic field values ($\zeta_B \ll 1$) escape losses dominate: 
convection for electrons with $E_\mathrm{e} \lesssim 1$~TeV, and diffusion for $E_\mathrm{e} \gtrsim 1$~TeV.
For protons, on the other hand, the cooling processes taken into account are proton-proton inelastic collisions (p-p)
and adiabatic expansion. However, protons do not suffer significant energy losses (Fig.~\ref{fig:tiempos}) and escape completely dominates: convection for protons with $E_\mathrm{p} \lesssim 1$~TeV, and diffusion for 
$E_\mathrm{p} \gtrsim 1$~TeV.

\begin{figure}
    \resizebox{\hsize}{!}{\includegraphics[width=0.35\textwidth, angle=270]{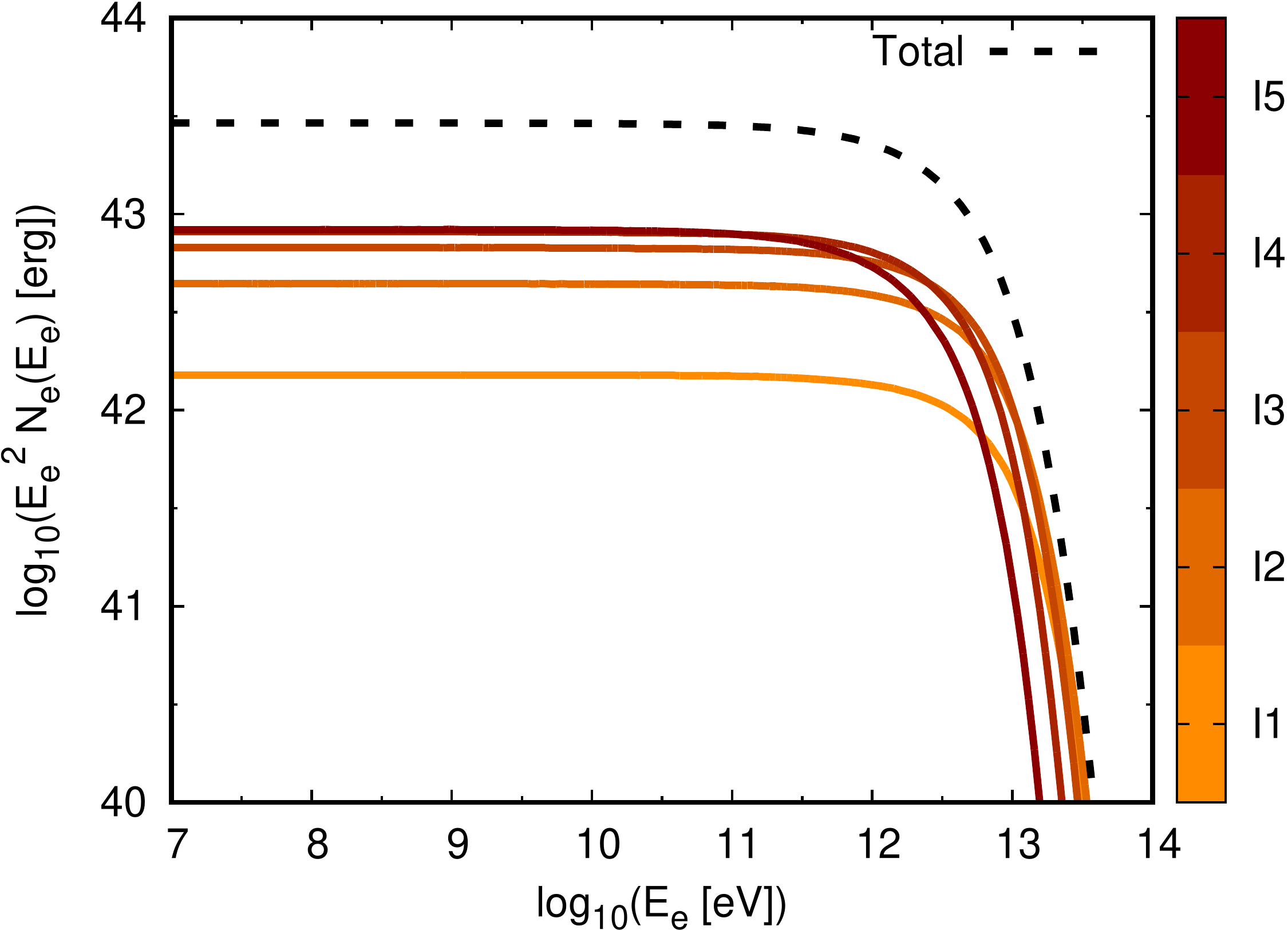} }    
    \resizebox{\hsize}{!}{\includegraphics[width=0.35\textwidth, angle=270]{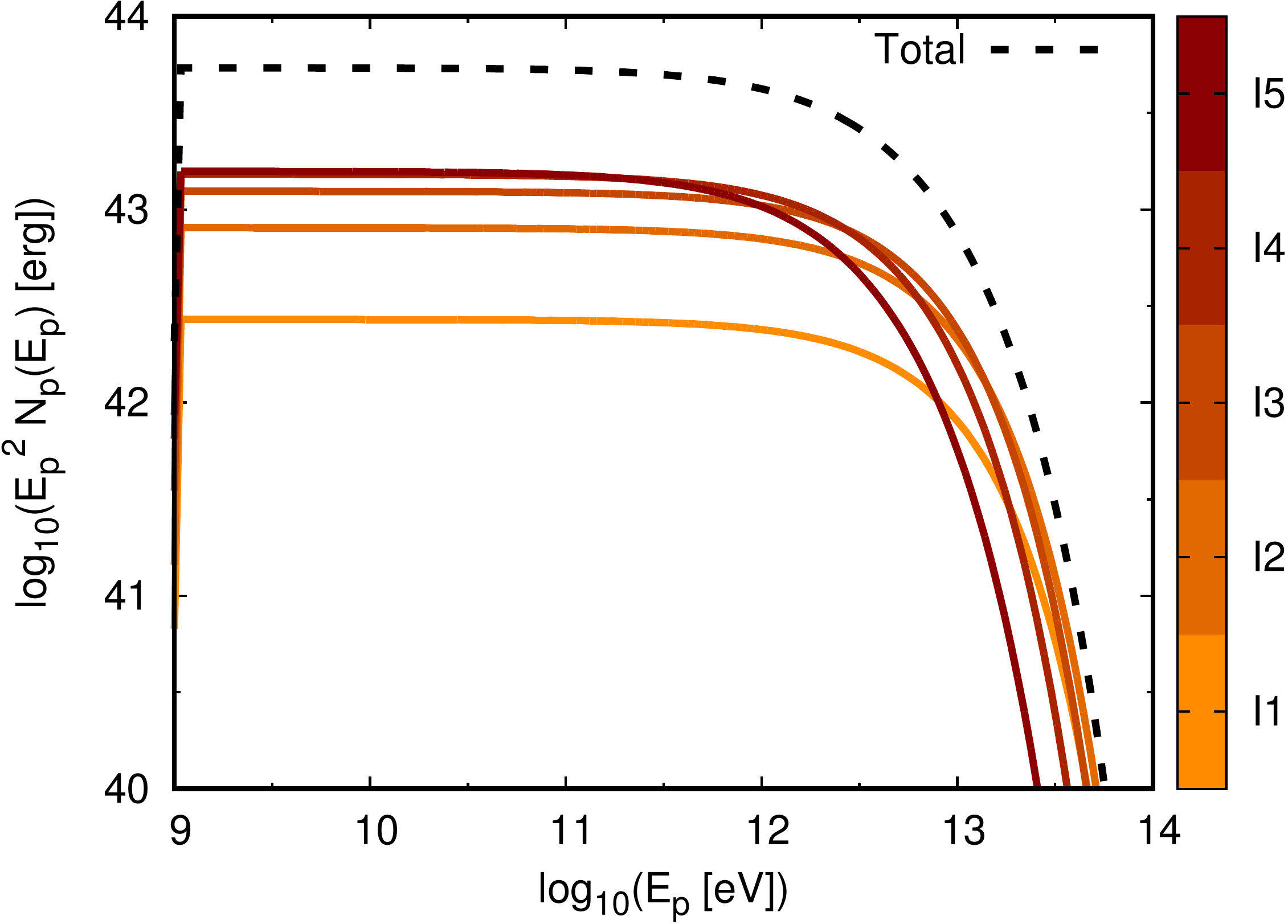} }   
    \caption{Electron (top) and proton (bottom) energy distribution for the generic case presented in 
    Table~\ref{table:parameters}. The colour scale represents five different sections (I1--I5) of the emitter 
    that correspond to intervals of length $\Delta \theta = 0.15 \, \pi$, starting in $\theta \in [0,0.15\,\pi)$ 
    for I1, and so on. The black dashed line is the total particle distribution (i.e. the sum of all curves).}
    \label{fig:dist} 
  \end{figure}

When convection dominates the loss time, particles move along the BS with an energy distribution that
keeps the same spectral index as the injected distribution. Adiabatic losses slightly soften the particle energy
distribution as the particles stream along the emitter. Additionally, for high-energy electrons with 
$E_\mathrm{e} \gtrsim 1$~TeV a combination of synchrotron and IC cooling in the Thomson regime can also contribute to a 
spectral softening. Figure~\ref{fig:dist} shows the described behaviour for different linear emitters, as well as the
total electron distribution and proton distribution at the BS, also calculated for the generic scenario case presented in
Sect.~\ref{sec:one_vs_multi}.

The IC cooling timescale for each photon field was calculated using the formulae given by \citet{Khangulyan2014}, suitable for black-body-like spectra. For the case of the stellar UV photon field, we consider that the star emits as a black body with a temperature $T_\star \sim 40\,000$~K and a dilution factor
$\kappa_\star = \left[ R_\star/(2 R(\theta)) \right]^2$, since $R(\theta)$ is the distance from the star to the BS at the position $\theta$. The dilution factor is defined by \citet{Khangulyan2014} as the ratio of radiation energy density in the emitter to radiation energy density within a thermal gas. The interaction angle for the scattering process is calculated as the angle between the direction of motion of the emitting electron (which is the line of sight, as the emission is beamed in the direction of the emitting electron) and the radial direction of the stellar photon; naturally, this angle varies with $\theta$. For the case of the IR photon field produced by the dust, we assume isotropy within the NT emitter. Its spectrum is well-approximated with a Planck law of temperature $T_\mathrm{IR} \sim 100$~K \citep{Draine2011, Kobulnicky2017}. The observational evidence shows that the emitting dust usually surrounds the cavity produced by the shocked SW \citep{Kobulnicky2017}. In consequence, if the dust were optically thick, it would be appropriate to set $\kappa_\mathrm{IR}=1$, which was adopted by \citet{delValle2012}. However, the dust is not strictly optically thick.\footnote{The assumption that the dust emits as a black body leads to an overestimation by several orders of magnitude of the observed IR emission, as $\sigma {T^4}_\mathrm{IR}{R_0}^2 \gg L_\mathrm{IR}$.}
This issue has been addressed by \citet{DeBecker2017} through the introduction of a ``normalization factor'' (also known as a grey body). This is equivalent to setting a proper ``dilution factor'' in the formalism given by \citet{Khangulyan2014}. Considering $U_\mathrm{BB} = 4 \,(\sigma/c)\, T_\mathrm{IR}^4$, the dilution factor of the IR photon field along the BS, $\kappa_\mathrm{IR}(\theta) = U_\mathrm{IR}(\theta)/U_\mathrm{BB}$, can be approximated as
\begin{equation}
 \kappa_\mathrm{IR}(\theta) \approx \frac{L_\mathrm{IR}}{4 \pi \sigma T_\mathrm{IR}^4 R(\theta)^2},
\end{equation}
where we have considered $U_\mathrm{IR} \approx  L_\mathrm{IR}/[\pi \, R(\theta)^2 \, c]$. As the extended IR radiation is produced in a region of size $\sim R_0$ surrounding the NT emitter, the above expression for $U_\mathrm{IR}$ should be valid, at least, for a region of size $\sim R_0$ centred in the apex, and therefore for the brightest portion of the NT emitter. We note that far away from a point-like source (such as the star) the energy density is $U = L/(4\pi r^2\,c)$, so in the more external regions of the BS we probably overestimate this value. However, if the plasma is not completely optically thin ($\tau \lesssim 1$), part of the more internal emission should be reprocessed and isotropised, so we expect that the above expressions serve as a decent approximation even in the outer regions of the NT emitter as well. Moreover, the election of $R(\theta)$ in the denominator instead of simply $R_0$ is a phenomenological (not particularly physically motivated) approach to reproduce the observed decay of the IR brightness away from the apex of the BS.\footnote{With this prescription the energy density at a position $\theta$ in the BS is a factor $(R_0/R(\theta))^2$ smaller than at the apex.} We emphasise that the detailed modelling of the (extended) IR field produces a negligible impact in the NT emission considering that it is produced mostly at distances from the apex $\lesssim R_0$ (Sect. 2.1). As we show in Sect.~\ref{sec:results}, the inclusion of a dilution factor for the IR photon field is enough to account for the incompatibility between some of the previous predictions and the upper limits set by recent observations in the X-ray and $\gamma$-ray energy bands. We note that even though the energy density of the stellar UV field exceeds the energy density of the dust IR field, cooling due to IC-IR dominates over IC-star for electrons with energies $E_\mathrm{e} \gtrsim 200$~GeV. This happens because, for the latter, IC takes place much deeper in the Klein-Nishina regime.  

The normalisation of the evolved particle distribution depends on the power injected 
perpendicularly to the shock surface, $L_{\mathrm{w},\perp} \approx 50$\% of the total wind power,
and on the fraction of that luminosity that goes to
NT particles, $f_\mathrm{NT}$, which is a free parameter of the model. Given that there is no tight constraint on how
the energy is distributed in electrons and protons, we consider two independent parameters 
$f_\mathrm{NT,e}$ and $f_\mathrm{NT,p}$. It is worth noting that ISM-termination shocks of 
massive stars may transfer up to a $\sim 10\%$ of their energy into relativistic protons 
\citep{Aharonian2018}.


 \subsection{Numerical treatment} \label{sec:numerical}
  
 We apply the following procedure to obtain the NT particle distribution:
 
 \begin{enumerate}
 
  \item{We obtain the location of the CD in the $XY$-plane in discrete points $\left( x_i(\theta_i),y_i(\theta_i) \right)$. 
  We characterise the position of the particles in their trajectories along the BS region through 1D cells 
  (or linear-emitter segments) located at those points.}
  
  \item{We compute the thermodynamic variables at each position $i$ in the trajectory (Sect.~\ref{subsec:hydrodynamics}).}
  
  \item{The wind fluid elements reach the RS at different locations, from where they are convected along the BS. We simulate the different trajectories by taking different values of $i_\mathrm{min}$: the case in which $i_\mathrm{min} = 1$ 
  corresponds to a line starting in the apex of the BS, whereas $i_\mathrm{min} = 2$ corresponds to a line that starts a 
  bit further in the BS, and so on (Fig.~\ref{fig:lines}). 
  The axisymmetry allows us to compute the trajectories only for the 1D emitters with $y \geq 0$.}
  
  \begin{figure}
    \centering
    \includegraphics[width=0.45\textwidth]{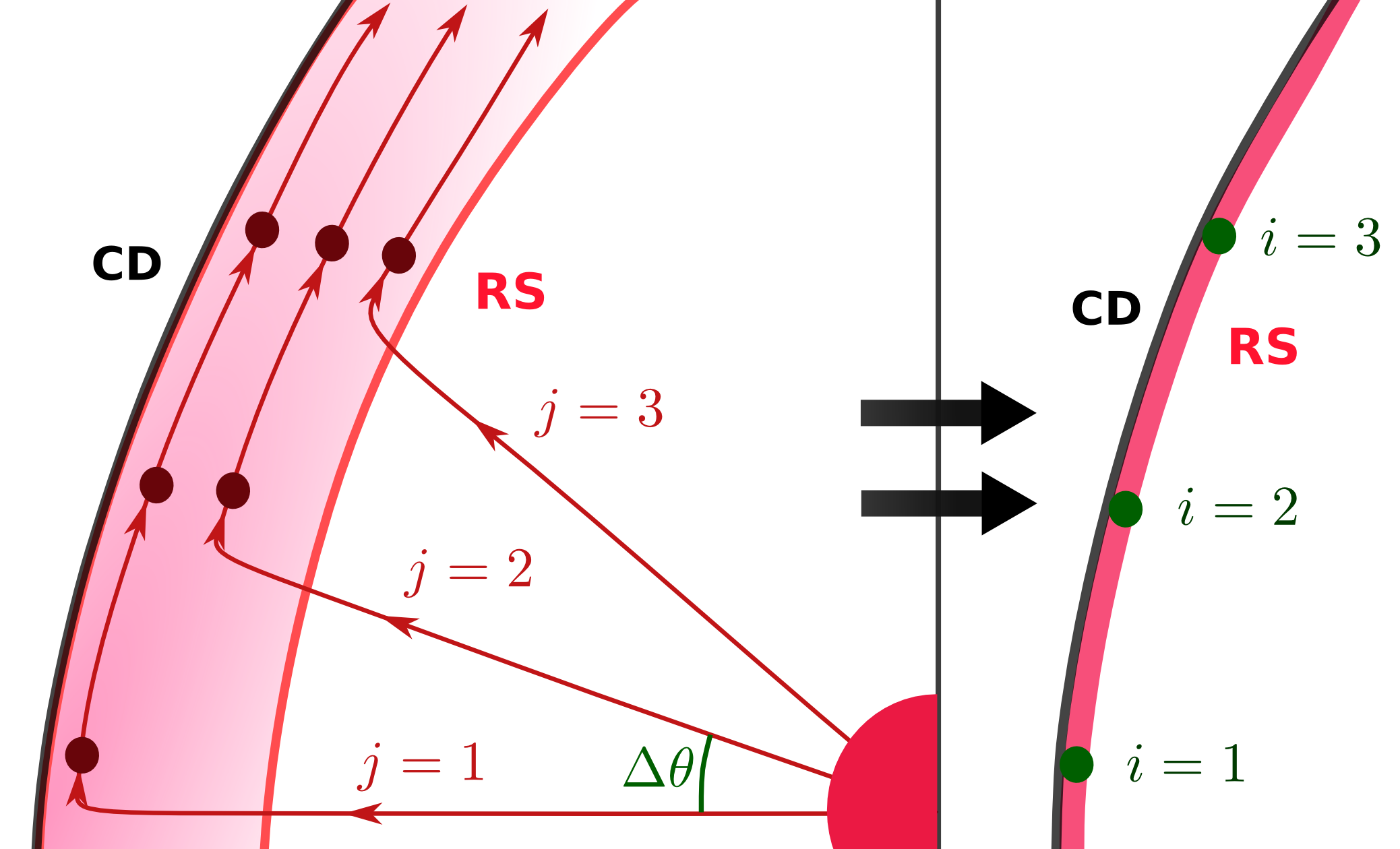} 
    \caption{Illustration of the model for the spatial distribution of NT particles in the BS region. On the left side we show different linear emitters, named as $j=1,2,3$ according to the incoming stellar wind fluid line. On the right side we represent three cells of a bunch of linear emitters, obtained from summing at each location the contributions from the different linear emitters (see Sect.~\ref{sec:numerical}).}
    \label{fig:lines}
  \end{figure}

  \item{We calculate the power available to accelerate NT particles (either protons or electrons) at each position $i$ as 
  
  \begin{equation} \label{eq:L_inj}
     \Delta L_\mathrm{NT}(\theta_i) = f_\mathrm{NT}\;L_{\mathrm{w},\perp}(\theta_i)\;\frac{\Delta \Omega(\theta_i)}{4 \pi},
  \end{equation}
  where $L_{\mathrm{w},\perp} = 0.5\;\dot{M}\;{v^2_{\mathrm{w},\perp}}$, and 
  $\Delta \Omega = \sin \theta\; \Delta \theta\; \Delta \phi$ is the solid angle subtended by the cell. The quantity 
  $\Delta L_\mathrm{NT}$ can be considered as a discrete version of a differential luminosity per surface element.}
    
  \item{Relativistic particles are injected in only one cell per linear emitter (that corresponding to $i_\mathrm{min}$),
  and each linear emitter is independent from the rest. For a given linear emitter, particles are injected at a location
  $\theta_i$ with an energy distribution 
  \mbox{$Q(E,\theta_i) = Q_0(\theta_i) E^{-p} \exp{(-E/E_\mathrm{cut}(\theta_i))}$}. The normalisation constant $Q_0$ is
  set by the condition \mbox{$\int E \, Q(E,\theta_i) \, \mathrm{d}E = \Delta L_\mathrm{NT}(\theta_i)$}, and the particle maximum energy 
  $E_\mathrm{cut}(\theta_i)$ by equating the acceleration time to the characteristic loss time, which takes into account
  both cooling and escape losses. The expressions used to calculate the different timescales are given below.
  
  The particle acceleration timescale is
  \begin{equation}\label{eq:t_ac}
    t_\mathrm{acc} = \eta_\mathrm{acc}(\theta) E_{e,p}/(B(\theta) \, c \, q) \; \mathrm{s},
   \end{equation}
  \noindent where the acceleration efficiency is $\eta_\mathrm{acc}(\theta) = 2 \pi E (c/v_\perp(\theta))^2$. 
  The characteristic escape timescale of the particles is given by
  \begin{align}\label{eq:t_esc}
    t_\mathrm{esc} &= {\left(t_\mathrm{conv}^{-1} + t_\mathrm{diff}^{-1} \right)}^{-1} \, \mathrm{s}\\
    t_\mathrm{conv} &= R(\theta) / v_\parallel(\theta) \, \mathrm{s} \\
    t_\mathrm{diff} &= H(\theta)^2/(2 D_\mathrm{Bohm}) \, \mathrm{s},
   \end{align}  
   \noindent where we consider a characteristic convection timescale, and diffusion in the Bohm regime such that 
   $D_\mathrm{Bohm} = r_\mathrm{g}c/3$, where $r_\mathrm{g} = E/(q\,B)$ is the gyroradius of the particle. 
   
   Denoting $n_\mathrm{ssw}$ the particle number density in the shocked SW, the proton and electron energy losses are
   \begin{align}\label{eq:t_pp}
    t_\mathrm{pp} &= 10^{15} / n_\mathrm{ssw}(\theta) \, \mathrm{s} \\
    t_\mathrm{adi} &= \frac{3}{v_\parallel(\theta)} \frac{\mathrm{d}s(\theta)}{\mathrm{d}(-\log{\rho(\theta)})} \, \mathrm{s}\\
    t_\mathrm{cool,p} &= {\left(t_\mathrm{pp}^{-1} + t_\mathrm{adi}^{-1} \right)}^{-1} \, \mathrm{s}
   \end{align} 
   and
   \begin{align}\label{eq:t_rad}
    t_\mathrm{br} &= 10^{15} / n_\mathrm{ssw}(\theta) \, \mathrm{s} \\
    t_\mathrm{sy} &= \left[ 1.6 \times 10^{-3} B_\parallel(\theta)^2 E_e \right]^{-1} \, \mathrm{s} \\
    t_\mathrm{IC} &= {\left(t_\mathrm{IC,\star}^{-1} + t_\mathrm{IC,IR}^{-1} \right)}^{-1} \, \mathrm{s} \\
    t_\mathrm{cool,e} &= {\left(t_\mathrm{br}^{-1} + t_\mathrm{sy}^{-1} + t_\mathrm{IC}^{-1} + t_\mathrm{adi}^{-1} \right)}^{-1} \, \mathrm{s},
   \end{align}  
  respectively.}

   \item{At the injection cell, the steady-state particle distribution is approximated as 
   \mbox{$N_0(E,i_\mathrm{min}) \approx Q(E) \times \min{\left(t_\mathrm{cell},t_\mathrm{cool} \right)}$}, where
   \mbox{$t_\mathrm{cell} = s_\mathrm{cell}(\theta)/v_\parallel(\theta)$} is the cell convection time (i.e. the time particles spend in each cell).

  By the time the relativistic particles reach the next cell in their trajectory, their energy has diminished from $E$ to $E'$, but the 
  total number of particles must be conserved so that $N(E) \, \mathrm{d}E = N(E') \, \mathrm{d}E'$. From that condition we 
  can obtain the evolved version of the injected distribution from
  \begin{equation} \label{eq:Ne_i}
   N(E',i+1) = N(E,i)\frac{\lvert \dot{E}(E,i)\rvert}{\lvert \dot{E}(E,i+1)\rvert},
  \end{equation}
  where $\dot{E}(E,i) = E/t_\mathrm{cool}(E,i)$ is the cooling rate for particles of energy $E$ at the position $\theta_i$. The energy $E'$ is
  given by the condition $t_\mathrm{cell} = \int_{E}^{E'} \dot{E}(\tilde{E},i) \mathrm{d}\tilde{E}$.}
  
  \item{We repeat the same procedure varying $i_\mathrm{min}$, which represents different linear emitters in the $XY$-plane.}
  
  \item{We obtain the total steady-state particle energy distribution at each cell, $N_\mathrm{tot}(E,i)$, by summing the 
  distributions $N(E,i)$ obtained for each linear emitter. The result is what we call  a bunch of linear emitters, which represents a wedge of width $\Delta \phi$ from the total BS structure (calculated at a fixed $\phi$).}
  
  \item{We can obtain the evolved particle energy distribution for each position $(x_i,y_i,z_i)$ of the BS surface in a 3D geometry. We achieve this by distributing -- via a rotation -- the bunch
  of linear emitters around the axis given by the direction of motion of the star. By
  doing so, we end up with many bunches of linear emitters, each with a different value of the azimuthal
  angle $\phi$. The particle energy distribution is the same in all the bunches because of the azimuthal
  symmetry. We note that the normalisation of the particle energy distribution already takes into account
  the total number of bunches, as for $m_\phi$ bunches we take $\Delta \phi = 2\pi / m_\phi$ in 
  Eq.~\ref{eq:L_inj}.}
   
  \end{enumerate}
  

\subsection{Non-thermal emission}\label{sec:nt_emission}

 Once the distribution of particles at each cell $(x_i,y_i,z_i)$ is known, it is possible to calculate the emission at 
 each cell by the previously mentioned radiative processes, and also to obtain the total emission from the modelled region. 
 Radio, X-ray, and $\gamma$-ray absorption processes are not relevant given the conditions (size, density, and target 
 fields) of the BS region. The relevant radiative processes are: IC \citep{Khangulyan2014}, synchrotron, p-p, and relativistic Bremsstrahlung
\citep[see e.g.][and references therein]{Bosch-Ramon2009}. Except for the IC with the stellar UV photons, which depends on
the star-emitter-observer geometry, the other radiative processes can be regarded as isotropic, given that the NT particle
population is isotropic due to an irregular magnetic field component in the shocked gas. 
Nevertheless, the presence of an orderly $B$-component leads to some degree of anisotropy in the synchrotron emission. After calculating the emission at
each cell, we can take into account absorption effects along the radiation path if necessary (e.g. free-free absorption of
radio photons in the unshocked stellar wind).

 \subsection{Synthetic radio-emission maps}\label{sec:synthetic_maps}

The total spectral energy distribution (SED) is obtained as the sum of the emission from all the bunches of 
linear emitters. This SED does not contain all the information available from the model, which can be contrasted with data
from spatially resolved radio observations. Therefore, synthetic radio emission maps are valuable complementary
information of the morphology predicted by our model, which can help to interpret observations such as the ones reported
by \citet{Benaglia2010}. 

\begin{figure*}
    \centering
    \includegraphics[width=0.48\textwidth, angle=270]{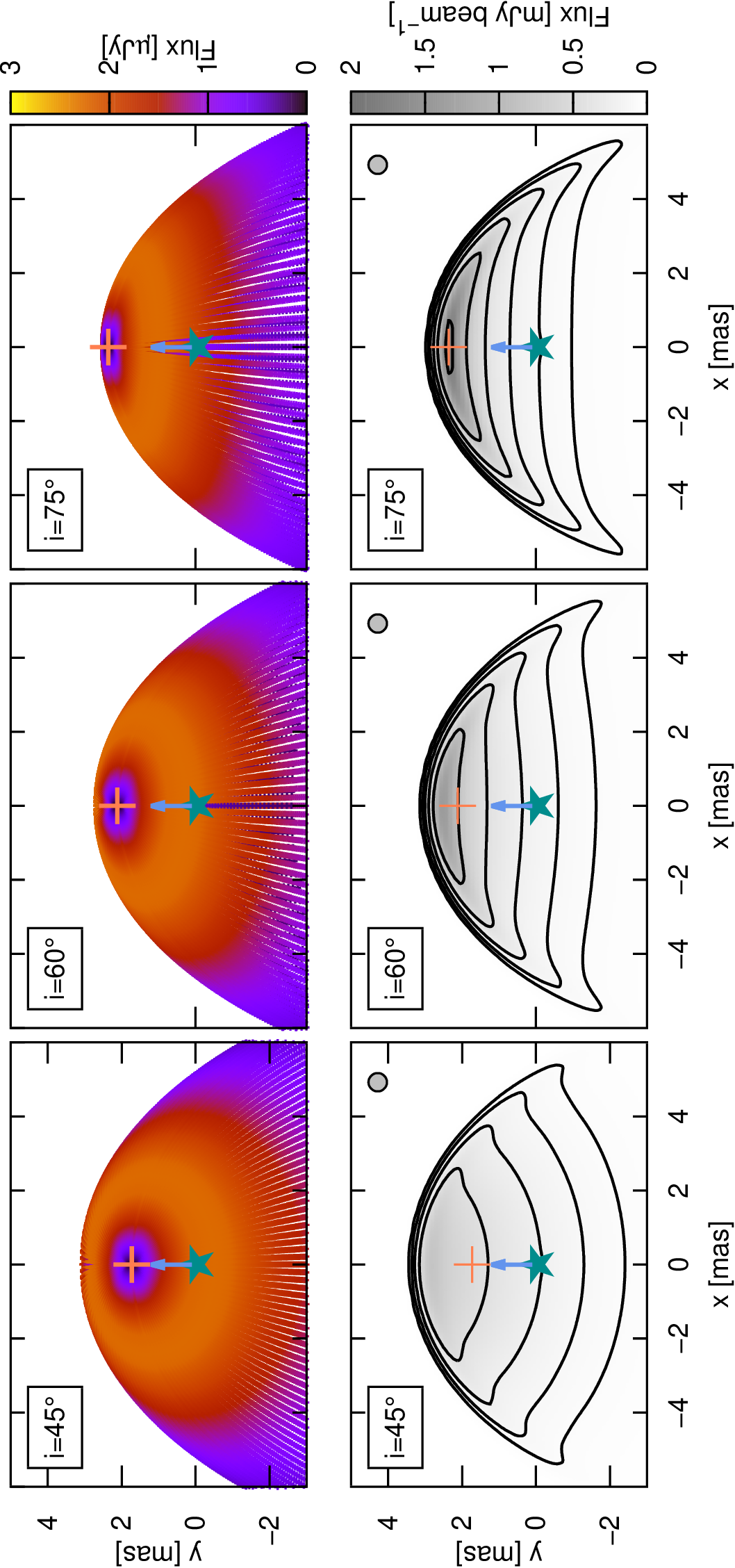} 
    \caption[]{Projected emission maps before (top) and after (bottom) convolution with a Gaussian beam 
    with $\sigma_x = \sigma_y = 12\arcsec$, calculated for the parameters of the generic scenario given in
    Table~\ref{table:parameters} but for different inclinations $i$. The 
    pink cross marks the position of the projected stagnation point, $R_0 \times \sin{i}$. The contours in the bottom
    plots are at 0.05, 0.1, 0.2, 0.4, 0.8, and 1.6~mJy~beam$^{-1}$.} 
    \label{fig:radio_maps_generic}
  \end{figure*}

To produce synthetic radio maps at a given frequency, we first project the 3D emitting structure in the plane of the sky, obtaining a 2D distribution of flux (Fig.~\ref{fig:radio_maps_generic}). We then cover this plane adjusting at each location of the map an elliptic Gaussian that simulates the synthesised (clean) beam. 
If the observational synthesised beam has an angular size $a \times b$, each Gaussian has $\sigma_x=a/\sqrt{8\log2}$ and $\sigma_y=b/\sqrt{8\log2}$. 
At each pointing we sum the emission from every location weighted by the distance between its projected position and the Gaussian centre: $\exp{ \left[ -(\Delta x^2/2\sigma_x^2) - (\Delta y^2/2\sigma_y^2) \right] }$. The result obtained is the
corresponding flux per beam at each position.

As shown in Fig.~\ref{fig:radio_maps_generic}, for observing angles $i \sim 90\degr$ the typical coma-shaped structure of the BS arises \citep{Peri2012, Kobulnicky2016}. On the other hand, for observing angles $i \lesssim 45\degr$ the BS shape is more circular\footnote{This is consistent with the statement by \citet{Kobulnicky2016} that BSs with $i < 65\degr$ 
are unlikely to be identified as such.} and, additionally, the emission gets very diluted spatially, so it would be difficult to detect such BSs. We also note that the position of $R_{0,\mathrm{proj}}$ lies closer to the star than the position of the maximum of emission; in fact, the maximum emission is always coincident with $R_0$ for values of $i > 45\degr$ (in angular units). 
Therefore, when measuring the value of $R_0$ from observational radio emission maps, a factor $\sin{i}$ should not be included.

 
\section{Results} \label{sec:results}


First, we compute the particle energy distribution and the NT emission 
for a generic scenario using a one-zone model approximation and the extended emission model. We then present analytical 
estimates of the NT emission dependence with the different system parameters. Finally, we apply our full emission model to the object \object{BD$+43^{\circ}3654$}.

\subsection{One-zone versus multi-zone model}\label{sec:one_vs_multi}

The basics of the one-zone model approximation are addressed by \citet{delValle2012}. Here we review only two aspects: 
the characteristic convection timescale and the IR photon field model. Previous one-zone models estimate the convection 
timescale as $t_\mathrm{conv} \sim H_0/v_\mathrm{w}$; however, the shocked fluid takes time to re-accelerate once it 
impacts on the stagnation point (Fig.~\ref{fig:termo_sw}). Moreover, the emitting area is more extended, similar to
$R_0$. We consider that a better estimate of the fluid convection time is $t_\mathrm{conv} \sim R_0/c_\mathrm{s}$, where
$c_\mathrm{s}$ is the sound speed in the shocked SW,
$c_\mathrm{s} \approx \sqrt{\gamma_\mathrm{ad} P/\rho} \approx v_\mathrm{w}/\sqrt{8}$. 
Regarding the modelling of the IR photon field, as discussed in Sect.~\ref{sec:model}, the assumption of a black-body-emitting surface embedding the emitter is not valid. We consider instead a radiation field approximated as a thermal 
(black-body-like) spectrum with an energy density of $U_\mathrm{IR} \approx L_\mathrm{IR}/(\pi {R_0}^2 c)$.

  \begin{figure}
    \resizebox{\hsize}{!}{\includegraphics[width=0.35\textwidth, angle=270]{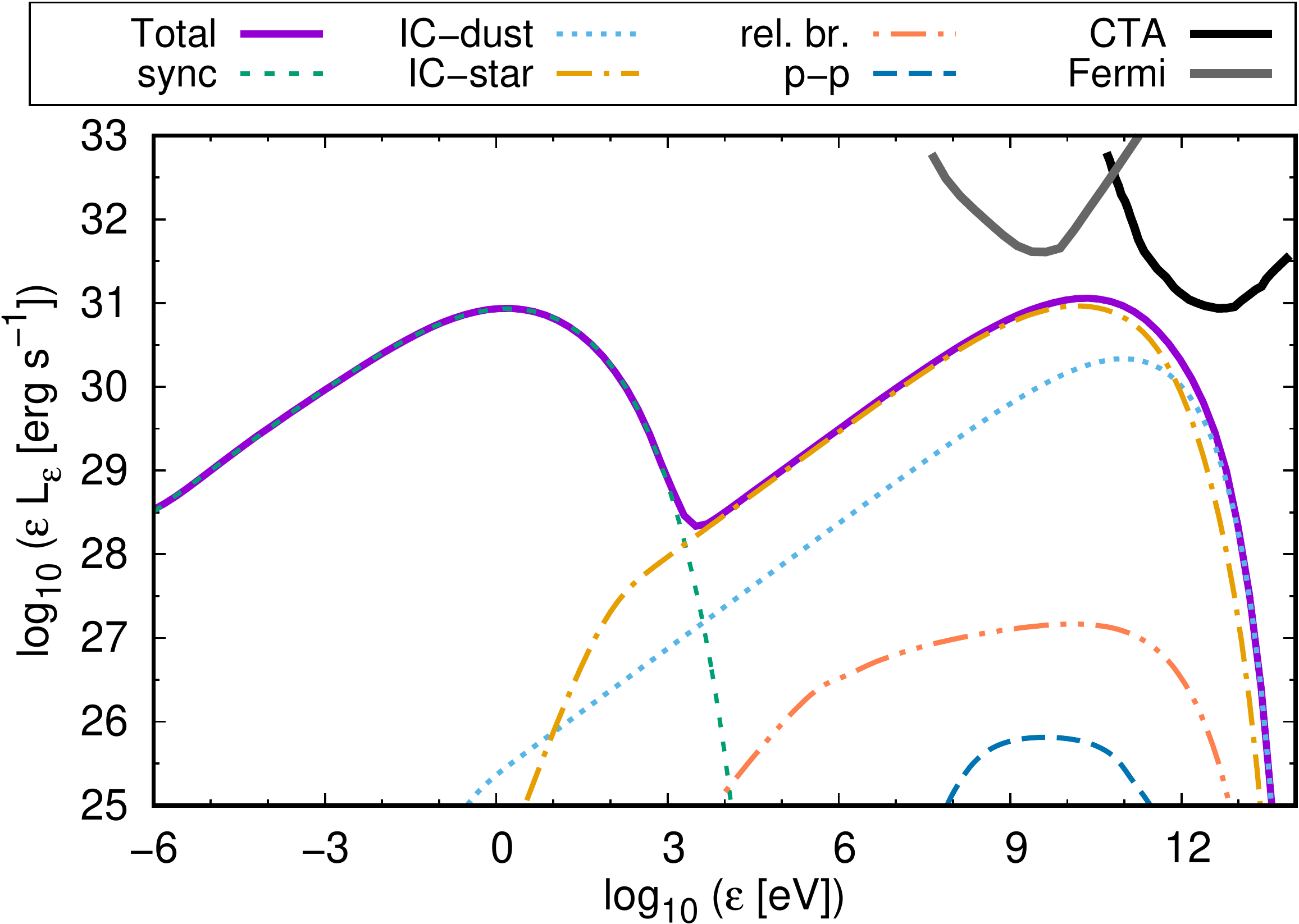}}   
    \resizebox{\hsize}{!}{\includegraphics[width=0.35\textwidth, angle=270]{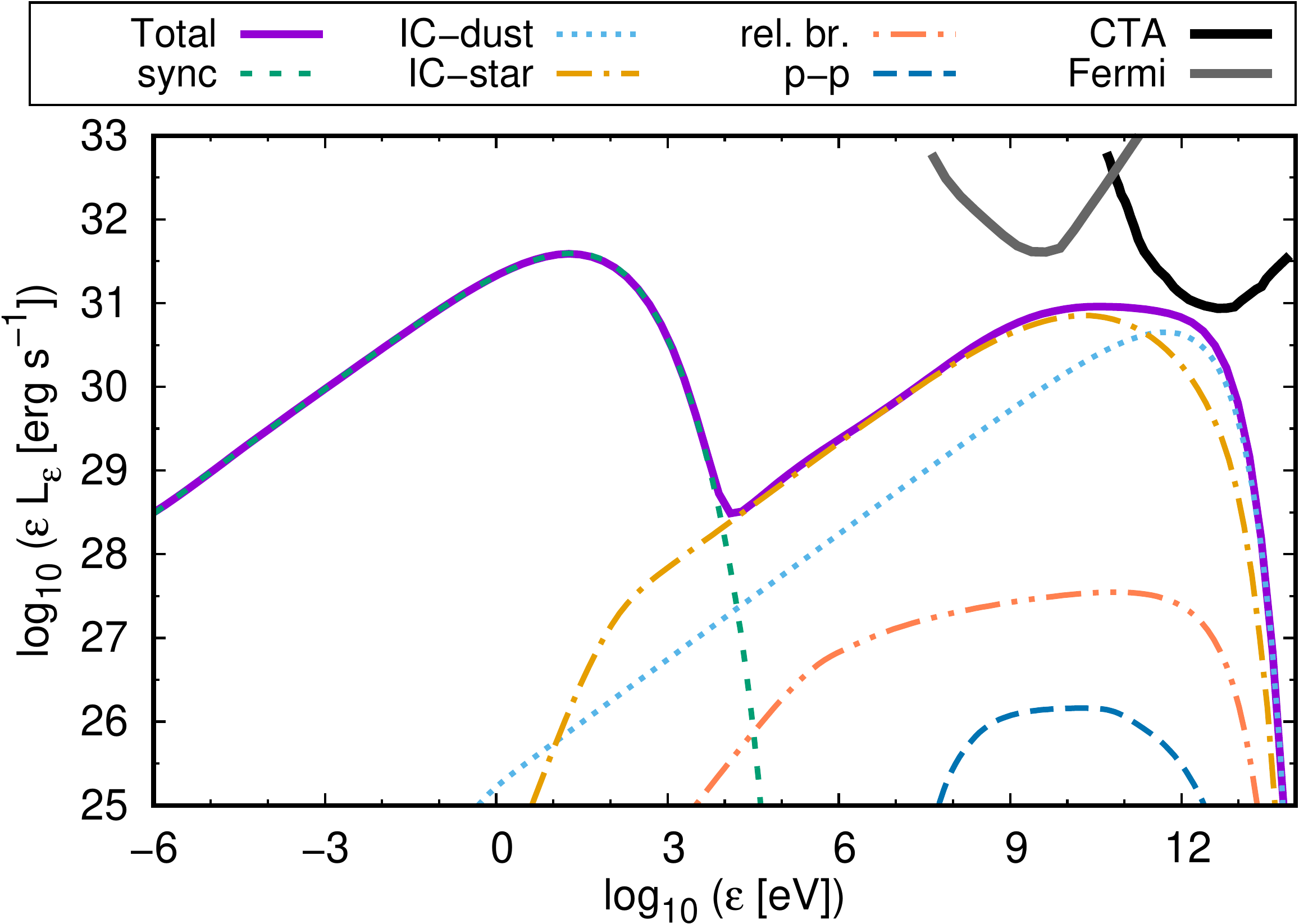}}  
    \caption[]{Comparison of the SEDs of the generic scenario with the parameters specified in
    Table~\ref{table:parameters} using a one-zone model (top) and a multi-zone model (bottom). The ten-year sensitivity curve of the \textit{Fermi} satellite is taken from \texttt{http://fermi.gsfc.nasa.gov}, and that of the 100-h CTA from \citet{Funk2013}.}
    \label{fig:sed_onezone_vs_extended}
  \end{figure}
  
  We assume that $10\%$ of the available injection luminosity goes into NT particles, equally distributed between 
  protons and electrons, that is, $f_\mathrm{NT,e} = f_\mathrm{NT,p} = 0.05$. We also fix the 
  magnetic field value adopting an intermediate value $\zeta_B = 0.1$, which yields $B \approx 20$~$\mu$G near 
  the apex of the BS; the full list of the selected parameters for this generic scenario is given in
  Table~\ref{table:parameters}. 
  
\begin{figure}
    \resizebox{\hsize}{!}{\includegraphics[width=0.35\textwidth, angle=270]{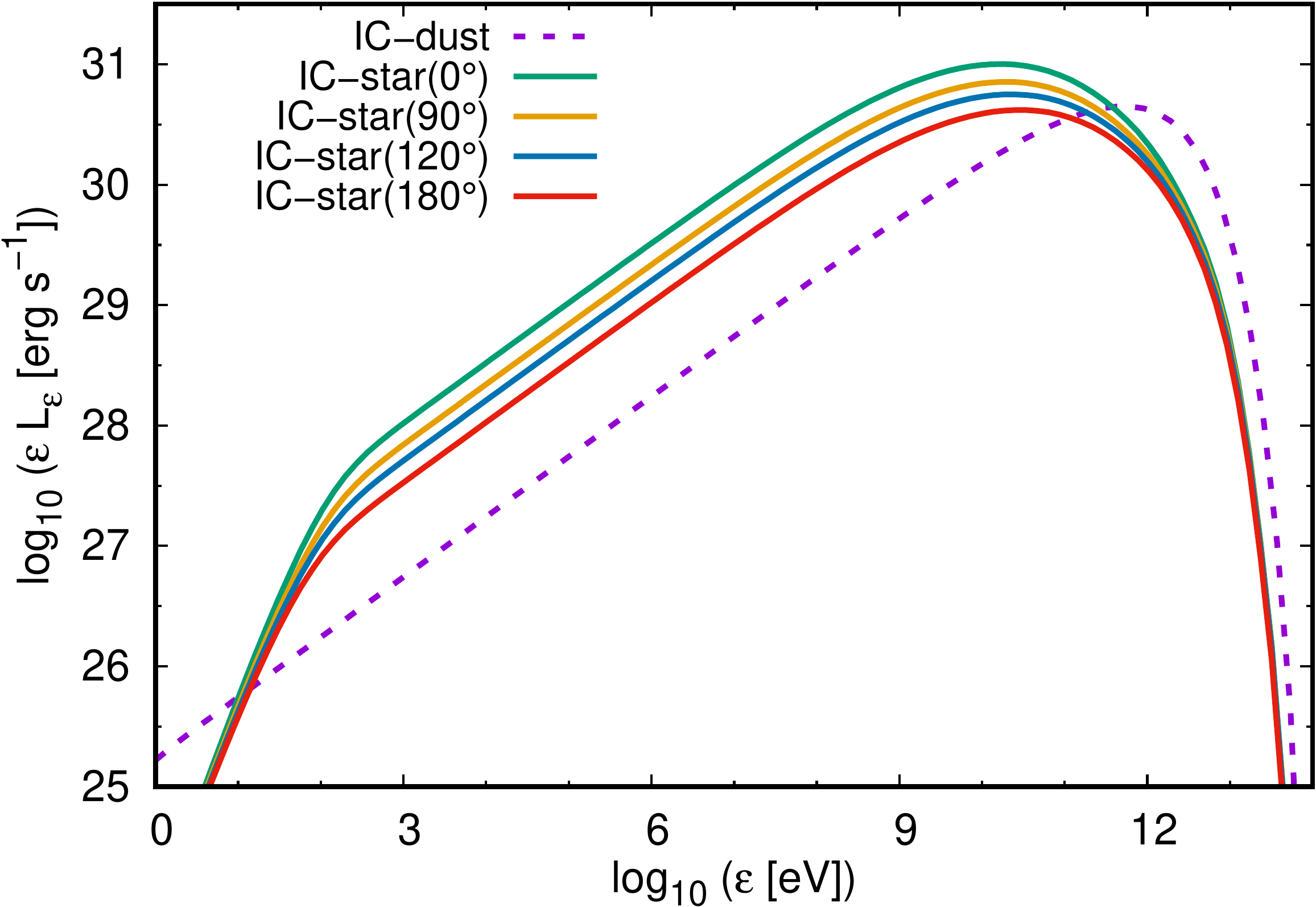}}  
    \caption[]{Comparison of $L_\mathrm{IC,star}$ for different values of the observing angle $i$. The simulations
    are calculated for the generic scenario with the parameters specified in Table~\ref{table:parameters}.}
    \label{fig:SEDs_IC_i}
  \end{figure}

  In Fig.~\ref{fig:sed_onezone_vs_extended} we show a comparison of the radiative outputs between the one-zone and 
  the multi-zone models. The one-zone model emission estimates agree with the extended model estimates within a factor
  of two to three, the largest discrepancies being found in the emission peaks related to the particle maximum energy, and in the 
  IC with the stellar UV photon field. As the anisotropic IC depends on the interaction angle and it varies along the 
  BS structure, we treated the IC-star as isotropic in the one-zone model, as using a single value for $i$ would be
  completely arbitrary and the impact that it has is quite large. For the extended model, using a value of $i = 90\degr$ is
  representative of the produced emission within a factor of two, as shown in Fig.~\ref{fig:SEDs_IC_i}.  
  
  \begin{table}
    \caption{Parameters of the generic system we modelled and the system \object{BD$+43^{\circ}3654$}.}
    \label{table:parameters}
    \centering
    \begin{tabular}{lccc}
    \hline \hline 
    \textbf{Parameter}                  &       \textbf{Generic}        &       \textbf{BD$+43^{\circ}3654$} & \textbf{Ref.} \\
    \hline
   $d$ [kpc]                            &       $1.0$                   &       $1.32$                  & 1 \\ 
   $i$                                  &       90$\degr$               &       $75\degr$               & - \\
   $R_{0,\mathrm{proj}}$ [']            &       -                       &         $3.2$                   & 2 \\  
   $L_\star$ [erg s$^{-1}$]             &       $2\times10^{39}$                &         $3.5\times10^{39}$      & 2 \\  
   $T_\star$ [K]                                &       $40\,000$               &       $40\,700$               & 1 \\
   $R_\star$ [$R_\sun$]                 &       $15.0$                  &       $19.0$                  & 1 \\   
   $\dot{M}_\star$ [$M_\sun$ yr$^{-1}$] &       $1\times 10^{-6}$       &       $9\times 10^{-6}$        & 3,4 \\  
   $v_{\infty}$ [km s$^{-1}$]           &       $2000$                  &       $2300$                  & 5,6 \\ 
    \cline{1-4}
   $v_\star$ [km s$^{-1}$]              &       $30$                    &       $40$                    & 6 \\
   $T_\mathrm{IR}$ [K]                  &       $100$                   &       $100$                   & 1 \\
   $n_\mathrm{ISM}$ [cm$^{-3}$]         &       $10$                    &       $15$                    & 2,6 \\
   $T_\mathrm{ISM}$ [K]                 &       $\sim 0$                &       $8000$                  & 2 \\
    \hline
   $L_{\mathrm{w},\perp}$ [erg s$^{-1}$]        &       $7\times10^{35}$                &         $8.9 \times 10^{36}$    & - \\  
   $f_\mathrm{NT,p}$                    &       $0.05$                  &       $0.5$                    & - \\
   $f_\mathrm{NT,e}$                    &       $0.05$                  &       $0.004$,$0.16$          & - \\
   $\zeta_B$                            &       $0.1$                   &       $0.01$,$1$              & 4 \\
   $p$                                  &       $2.0$                   &       $2.2$                   & - \\
    \hline
    \end{tabular} 
     \tablefoot{Regarding BD$+43^{\circ}3654$, the values adopted for various parameters are an intermediate between the values given by other authors. Details of the selection criteria can be found in the text.}
    \tablebib{(1) \citet{Kobulnicky2017}; (2) \citet{Kobulnicky2018}; (3) \citet{Peri2014}; (4) \citet{delValle2012};
    (5) \citet{Benaglia2010}; (6) \citet{Brookes2016}.}
  \end{table}
  
  For both models the available luminosity for NT particles in the BS is 
  \mbox{$L_{\mathrm{w},\perp} \approx 7 \times 10^{35}$~erg~s$^{-1}$}, from which only 
  \mbox{$L_\mathrm{NT} \approx 3 \times 10^{34}$~erg~s$^{-1}$} goes to each particle species.
  We show the radiated luminosities for both models in Table \ref{table:luminosities}, distinguishing the different contributions.
  Electrons radiate only $\sim 1$\% of their power, while for protons the radiated energy fraction is negligible. They escape as cosmic rays to the ISM.
    
    \begin{table}
    \caption{Luminosities produced by the different contributors of each model.}
    \label{table:luminosities}
    \centering
     \resizebox{\columnwidth}{!}{%
    \begin{tabular}{lcccc}
    \hline \hline 
    \textbf{Luminosity}                         &       \multicolumn{2}{c}{\textbf{One-zone model}} &       \multicolumn{2}{c}{\textbf{Extended model}} \\
    \hline
                                        & Value         & \% of $L_\mathrm{T}$ & Value & \% of $L_\mathrm{T}$ \\
    \hline
   $L_\mathrm{sy}$ [erg~s$^{-1}$]               &       $6.2\times 10^{31}$     & $44.5$  &       $2.5 \times 10^{32}$   & $81.8$ \\ 
   $L_\mathrm{Br}$ [erg~s$^{-1}$]               &       $1.5\times 10^{28}$ & $\ll 1$   &   $2.0\times 10^{28}$    & $\ll 1$        \\
   $L_\mathrm{IC,dust}$ [erg~s$^{-1}$]  &       $1.4\times 10^{31}$     & $9.9$   &       $2.7 \times 10^{31}$   & $9.0$  \\  
   $L_\mathrm{IC,\star}$ [erg~s$^{-1}$] &       $6.4\times 10^{31}$     & $45.6$  &       $2.8\times 10^{31}$    & $9.2$ \\  
   $L_\mathrm{pp}$ [erg~s$^{-1}$]               &       $8.0\times 10^{26}$     & $\ll 1$ &       $1.0\times10^{27}$     & $\ll 1$        \\
   \hline
   $L_\mathrm{T}$ [erg~s$^{-1}$]        & \multicolumn{2}{c}{$1.85\times 10^{32}$} & \multicolumn{2}{c}{$3.0\times 10^{32}$} \\
    \hline
    \end{tabular}}
  \end{table}

\subsection{Analytical estimates on emissivity scaling}\label{sec:analytical_estimates}

As we have shown in Sect.~\ref{sec:one_vs_multi}, the one-zone approximation gives a good estimate of the emission
obtained with a more complex model if one takes into account the modifications we propose. Thus, we can rely on the 
one-zone formalism to make simple estimates on how different system parameters impact on the NT radiative output.

Qualitatively, the NT radio luminosity depends on the ratio $t_\mathrm{conv}/t_\mathrm{sy}$, while the $\gamma$-ray
luminosity depends on the ratio $t_\mathrm{conv}/t_\mathrm{IC}$. IC-star dominates the SED for photon energies 
$\epsilon \lesssim 1$~GeV, while IC-IR dominates the SED above $\gtrsim 10$~GeV. In the X-ray energy band, synchrotron
dominates the NT emission, but a competitive or even dominant thermal contribution is possible\footnote{This statement 
is based in the usage of Eq.~24 from \citet{Christie2016} to estimate the thermal emission from the BS (not shown).
However, the simple HD model we use for the BS is not reliable for the calculation of its thermal contribution.}.

Quantitatively, on the one hand, 
$t_\mathrm{conv} \sim R_0/c_\mathrm{s} \propto \dot{M}^{0.5}\, v_\mathrm{w}^{-0.5}\, n_\mathrm{ISM}^{-0.5}\, v_\star^{-1}$
and $t_\mathrm{sy} \propto B^{-2} \propto \left( \dot{M}\, v_\mathrm{w}\, R_0^{-2} \right)^{-1} \propto n_\mathrm{ISM}^{-1}\,v_\star^{-2}$.
Thus, $L_\mathrm{sy} \sim L_\mathrm{NT,e} \times (t_\mathrm{conv}/t_\mathrm{sy}) 
\propto \dot{M}^{1.5}\,v_\mathrm{w}^{1.5}\,n_\mathrm{ISM}^{0.5}\,v_\star$, where we considered 
$L_\mathrm{NT,e} \propto \dot{M} \, v_\mathrm{w}^2$. As expected, high-velocity stars moving in a dense medium are good
candidates, although the most important quantities are intrinsic to the star, and are those related to the stellar wind: 
the denser and faster this wind is, the better are the chances that its BS will be detectable at radio frequencies.
In conclusion, in the search for synchrotron-emitting stellar BSs, it is more important to take into account the 
individual properties of the star than its runaway conditions. We note that the above scaling is not entirely valid for
electrons with $E_\mathrm{e} > 100$~GeV if $\zeta_B\sim 1$, as in that case synchrotron losses might dominate over convection
losses close to the apex, depending on the system parameters.

On the other hand, the $\gamma$-ray emission at energies above 10~GeV is dominated by IC with the dust IR photon field as already shown by \citet{delValle2012}. 
Assuming that $L_\mathrm{IR} \propto L_\star\,n_\mathrm{ISM}$ and $U_\mathrm{IR} \propto L_\mathrm{IR} R_0^{-2}$, which 
seems coherent and also has some empirical support \citep[][although with a high dispersion]{Kobulnicky2017},
we have $L_\mathrm{IC,IR} \propto L_\mathrm{NT,e} \times (t_\mathrm{conv}/t_\mathrm{IC,IR}) 
\propto \dot{M}^{1.5}\,v_\mathrm{w}^{0.5}\,n_\mathrm{ISM}^{1.5}\,v_\star\,L_\star$. 
Similar to the synchrotron emission, the most decisive factor determiming $L_{\rm IC,IR}$ is the mass-loss rate of the stellar wind. The above scaling condition is valid if $t_\mathrm{IC} > t_\mathrm{conv}$ (Fig.~\ref{fig:tiempos}).
The case for IC cooling dominated by stellar UV photons is similar: 
$t_\mathrm{IC,\star}^{-1} \propto U_\star \propto L_\star R_0^{-2}$, so that
$L_\mathrm{IC,\star} \propto L_\mathrm{NT,e} (t_\mathrm{conv}/t_\mathrm{IC,\star}) 
\propto \dot{M}^{1.5}\,v_\mathrm{w}^{0.5} \,n_\mathrm{ISM}^{0.5}\,v_\star L_\star$. 

If we further assume that $L_\star \propto \dot{M}^{0.5} \, v_\mathrm{w}^{0.5}$ 
\citep[e.g. Fig.~10 from][]{Muijres2012}, then we obtain 
$L_\gamma \propto \dot{M}^{2} \, v_\mathrm{w} \, n_\mathrm{ISM}^{0.5} \,v_\star$. As the low-frequency radio emission is
purely of synchrotron origin, we get 
\begin{equation}\label{eq:ratio_radio_gamma}
\frac{L_\mathrm{radio}}{L_\gamma} \propto \frac{1}{n_\mathrm{ISM}}\, \left( \frac{v_\mathrm{w}}{\dot{M}} \right)^{0.5}\,,
\end{equation}
and therefore, the best radio emitting candidates are not necessarily the best $\gamma$-ray emitting candidates, 
as the former prefer faster over denser winds, and the latter have a stronger dependence on the ambient density. In both cases the dependence with $v_\star$ coincides.

Finally, we recall that electrons radiate only a small fraction ($\sim 1$\%) of their power in the spatial scales
considered in the model (Sect.~\ref{sec:one_vs_multi}), whereas protons escape almost with no energy losses. The
luminosity injected in both relativistic electrons and protons that escape the modelled BS region is 
\mbox{$L_\mathrm{NT} \approx 3 \times 10^{34}$~erg~s$^{-1}$}. These cosmic rays could cool once they reach a region further
downstream, where the BS structure becomes very turbulent or even closes due to the pressure of the ISM 
\citep[e.g.][]{Christie2016}. However, we do not expect the energy release in the back flow of the BS to be very
significant, given the lack of detection of NT emission from stellar BSs. The diffusion and emission from cosmic rays accelerated at stellar BSs of stars moving inside molecular clouds has been studied by \citet{delValle2014}, although in such environments the target nuclei are more suitable for efficient proton p-p and electron relativistic Bremsstrahlung. 
Another possibility to take into account is that the laminar approximation of the shocked flow could be relaxed
so that mixing between the shocked SW and the shocked ISM could occur. Given that the ambient density in the
shocked SW is $\sim 0.01$~cm$^{-3}$, whereas $n_\mathrm{ISM} \sim 1-10$~cm$^{-3}$, the relativistic Bremmstrahlung 
and p-p emission could be enhanced by two to three orders of magnitude if mixing is efficient 
\citep[see e.g.][and references therein]{Munar2013}. 

\subsection{Application of the model to \object{BD$+43^{\circ}3654$}}\label{sec:BD43}

The massive O4If star \object{BD$+43^{\circ}3654$}, located at a distance of $d=1.32$~kpc, has a strong and fast wind
that produces a stellar BS. This BS was first identified by \citet{Comeron2007} using infrared data, and it has an extension in the IR sky of 8'. This was the first stellar BS to be detected at radio wavelengths, by \citet{Benaglia2010}. In their work, \citet{Benaglia2010} presented Very Large Array (VLA) observations at 1.43 and 4.86~GHz
(Fig.~\ref{fig:radio_maps}, top), from which a mean negative spectral index $\alpha \approx -0.5$ (with 
$S_\nu \propto \nu^\alpha$) was obtained. 
Their finding was indicative of NT processes taking place at the BS, in particular relativistic particle acceleration 
(most likely due to DSA at the shock) and synchrotron emission. This conclusion triggered a series of
works on the modelling of the broadband BS emission \citep{delValle2012, delValle2014, Pereira2016} 
using a simplified one-zone approximation. In those works, the radio emission was assumed to be of synchrotron origin,
and the radio observations were used as an input to characterise the NT electron energy distribution; the predicted 
high-energy emission came from IC up-scattering of ambient IR photons by relativistic electrons. Here, we revisit the
predictions from the previous radiative one-zone numerical models by applying our more consistent multi-zone model.

The data at radio frequencies allow us to characterise the relativistic electron population, which in turn can be used 
to model the broadband SED. The NT emission from IR to soft X-rays is completely overcome by the thermal emission from
the star and/or the BS, so the SED of NT processes can only be tested in the high-energy (HE) domain, where the NT
processes also dominate the spectrum. The electrons that produce the observed synchrotron radiation also interact with
the ambient radiation fields, producing HE photons through IC scattering \citep{Benaglia2010}. The relevant HE processes
are anisotropic IC with the stellar radiation field and isotropic IC with the dust IR radiation field. 
The role of secondary pairs in the radiative output is not expected to be relevant \citep{delValle2012}.

The reported emission from \object{BD$+43^{\circ}3654$} by \citet{Benaglia2010} is consistent with a canonical NT
spectral index $\alpha = -0.5$, which corresponds to an injection index of $p = -2\alpha+1 = 2$. However, it is possible
that the observed emission is actually the sum of the synchrotron NT emission and a contribution of thermal emission,
either from the stellar wind or the BS. The thermal emission has a positive spectral index and, therefore, in order to
produce the observed spectrum with $\alpha = -0.5$, a softer (i.e. more negative) intrinsic NT spectral index would be
required. We note, however, that the thermal contribution at low frequencies is likely to be small so a big deviation from
$p=2$ is not expected. Furthermore, \citet{Toala2016} derived an upper limit to the X-ray flux in the 0.4--4~keV below
$3.6\times 10^{-14}$~erg~cm$^{-2}$~s$^{-1}$ using \textit{XMM-Newton} observations. That upper limit also seems to favour
a softer spectrum (Fig.~\ref{fig:SEDs}), although the reported value is model-dependent, and it was derived for a 
power-law photon spectra with index $\Gamma = 1.5$; such a spectrum would be appropriate for the synchrotron SED below
0.1~keV, where the slope is $\sim 0.5$, but not in the X-ray energy band according to our model.\footnote{Recall that in 
the SED $\epsilon^2 N_\mathrm{ph}(\epsilon)$ is plotted, so that the slope is $2 - \Gamma$.} According to all these remarks, we decided to adopt a softer injection index of $p=2.2$, although we note that this value is not tightly constrained.

Following \citet{Brookes2016}, we assume that the star is moving in the warm ISM, with a temperature 
$T_\mathrm{ISM} = 8000$~K. \citet{Christie2016} addressed the effects of the non-zero ambient thermal pressure in terms
of a parameter $\alpha_\mathrm{th} = P_\mathrm{th}/P_\mathrm{kin}$, which is 
$\sim kT_\mathrm{ISM}/(m_\mathrm{p} v_\mathrm{ISM}^2) \approx 4 \times 10^{-2}$ for this case. Thus, this introduces 
only a small correction in the geometry (for instance, $R_0 \propto (1+\alpha_\mathrm{th})^{-1}$), which we account for only for completeness. 

The system spatial velocity is not well determined. The velocity in the plane of the sky is $\approx 38.4$~km~s$^{-1}$ 
\citep[][and references therein]{Brookes2016}. \citet{Benaglia2010} used the radial velocity of $-66$~km~s$^{-1}$ as the 
radial velocity of the star with respect to the surrounding ISM, which might not be accurate when accounting for 
Galactic rotation. In fact, if this were the case, the inclination angle should be 
$i \approx \arctan{(38.4/66.2)} \approx 30\degr$, whereas the observed emission map favours a high inclination angle 
$i > 60\degr$, which is nearly edge-on. As the radial velocity is for certain negative, we can assume $i < 90\degr$. In
order to reproduce the observed emission maps, we choose $i \approx 75\degr$, which is consistent with 
$v_\star \approx 40$~km~s$^{-1}$. The angular separation to the apex of the BS is $R_{0,\mathrm{proj}} \approx 3.2'$
\citep{Kobulnicky2017}, although there is some uncertainty when obtaining $R_{0,\mathrm{proj}}$ from the extended
emission, as discussed in Sect.~\ref{sec:synthetic_maps} for the case of radio emission, and for instance
\citet{Peri2014} gives a value of $R_{0,\mathrm{proj}} = 3.5'$. In Table~\ref{table:parameters} we list the adopted 
system and model parameters.

The wind velocity, $v_\mathrm{w}$, is an important parameter in the model as it affects the energy 
budget of NT particles, the acceleration efficiency, the position of $R_0$, and the convection timescale. In order to
maintain a good agreement between our model and the observational constraints, we choose a value of 
\mbox{$v_\mathrm{w} \approx 2300$~km~s$^{-1}$}, like \citet{Peri2014}. We note that \citet{Kobulnicky2018} assume
a higher value of $v_\mathrm{w} = 3000$~km~s$^{-1}$, but, under the assumption of Bohm diffusion, that leads to a 
more efficient acceleration, which yields a synchrotron peak at X-rays well above the observational constraints 
(see Fig.~\ref{fig:SEDs}). Alternatively, it is possible to reconcile a higher $v_\mathrm{w}$ if Bohm diffusion is 
not a good approximation for the acceleration of particles at the BS, turning in practice $\eta_\mathrm{acc}$ into a 
free parameter in the model \citep[e.g.][]{DeBecker2017}.

For the unshocked ISM, we consider a molecular weight $\mu_\mathrm{ISM} = 1.37$ \citep[e.g.][]{Kobulnicky2018}, which 
is only relevant to determine $R_0$. The mean molecular weight in the shocked SW is needed in order to derive the total
number of targets for relativistic Bremsstrahlung and p-p interactions; considering that the material is completely
ionised, for typical abundances we get $\mu_\mathrm{ssw} = 0.62$. With the selected parameters, the equipartition magnetic field at the apex of the BS is $\sim 100$~$\mu$~G, somewhat
smaller than the value of $\sim 300$~$\mu$~G estimated by \citet{delValle2012}. Recall that the magnetic field drops
along the BS according to Fig.~\ref{fig:termo_sw}.
We are left with only a few free parameters: the value of $B$ in the shocked SW, the fraction $f_\mathrm{NT,e}$ of energy 
converted to NT electrons, and the inclination angle. The latter only affects the IC-star SED within a factor of approximately two, as discussed in Sect.~\ref{sec:one_vs_multi}. As we will show in what follows, it is possible to constrain these
parameters from the measured radio fluxes and morphological information from the resolved NT radio emission region. With 
this purpose we explore two extreme scenarios below.

\subsubsection{Scenario with low magnetic field} \label{subsec:lowB}

We apply the model described in Sect.~\ref{sec:model} to a scenario with a low magnetic field (i.e. $\zeta_B \ll 1$). 
To reproduce the radio observations from \citet{Benaglia2010}, we fix the values of the following parameters: $i=75\degr$, $f_\mathrm{NT,e} = 0.16$, $\zeta_B= 0.01$, which leads to $B \sim 10$~$\mu$G and 
$B_\star < 85$~G. We also fix a high value of $f_\mathrm{NT,p} = 0.5$ to obtain an upper limit to the p-p luminosity.
 
\begin{figure}
    \resizebox{\hsize}{!}{\includegraphics[width=0.35\textwidth, angle=270]{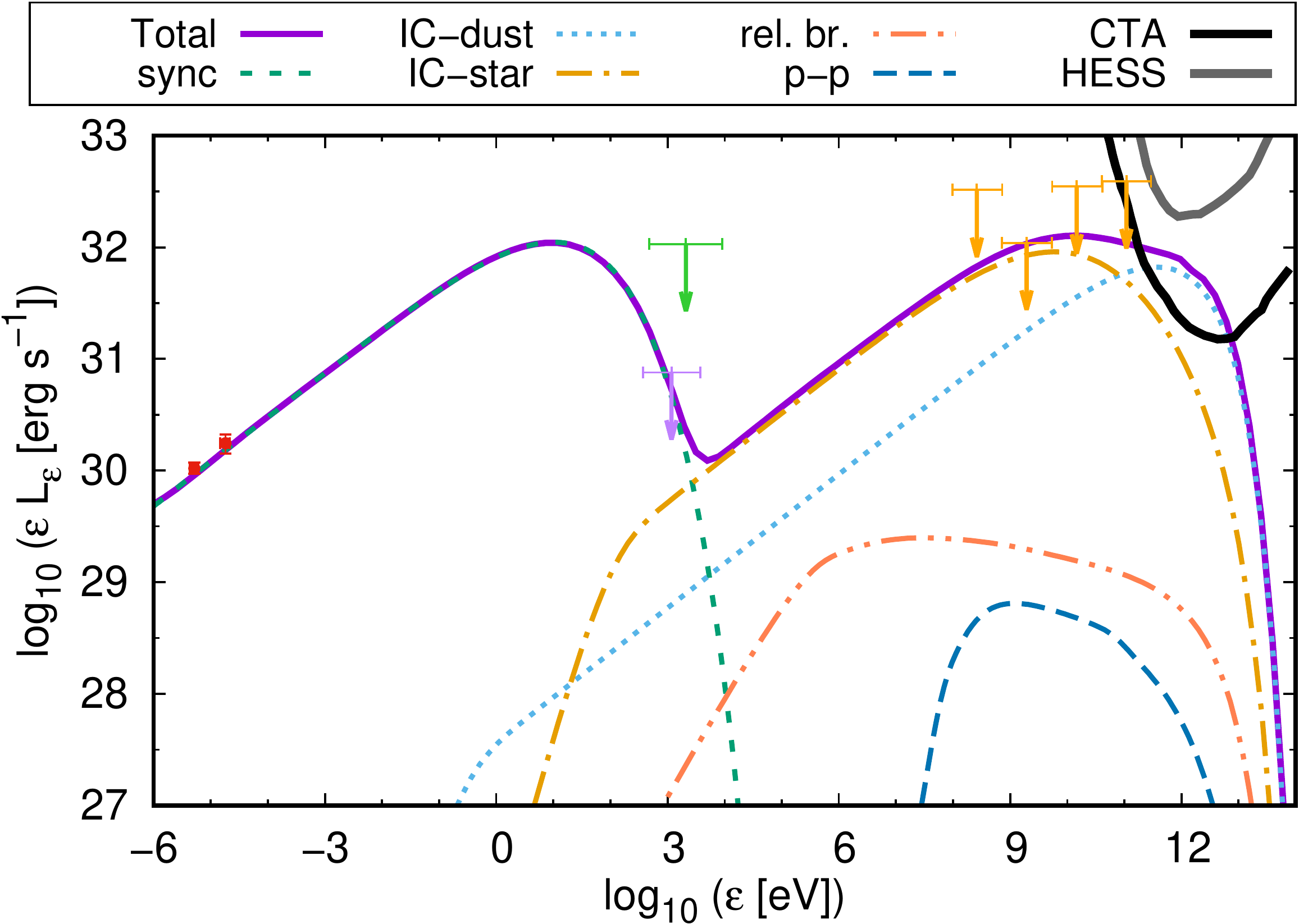}}   
    \resizebox{\hsize}{!}{\includegraphics[width=0.35\textwidth, angle=270]{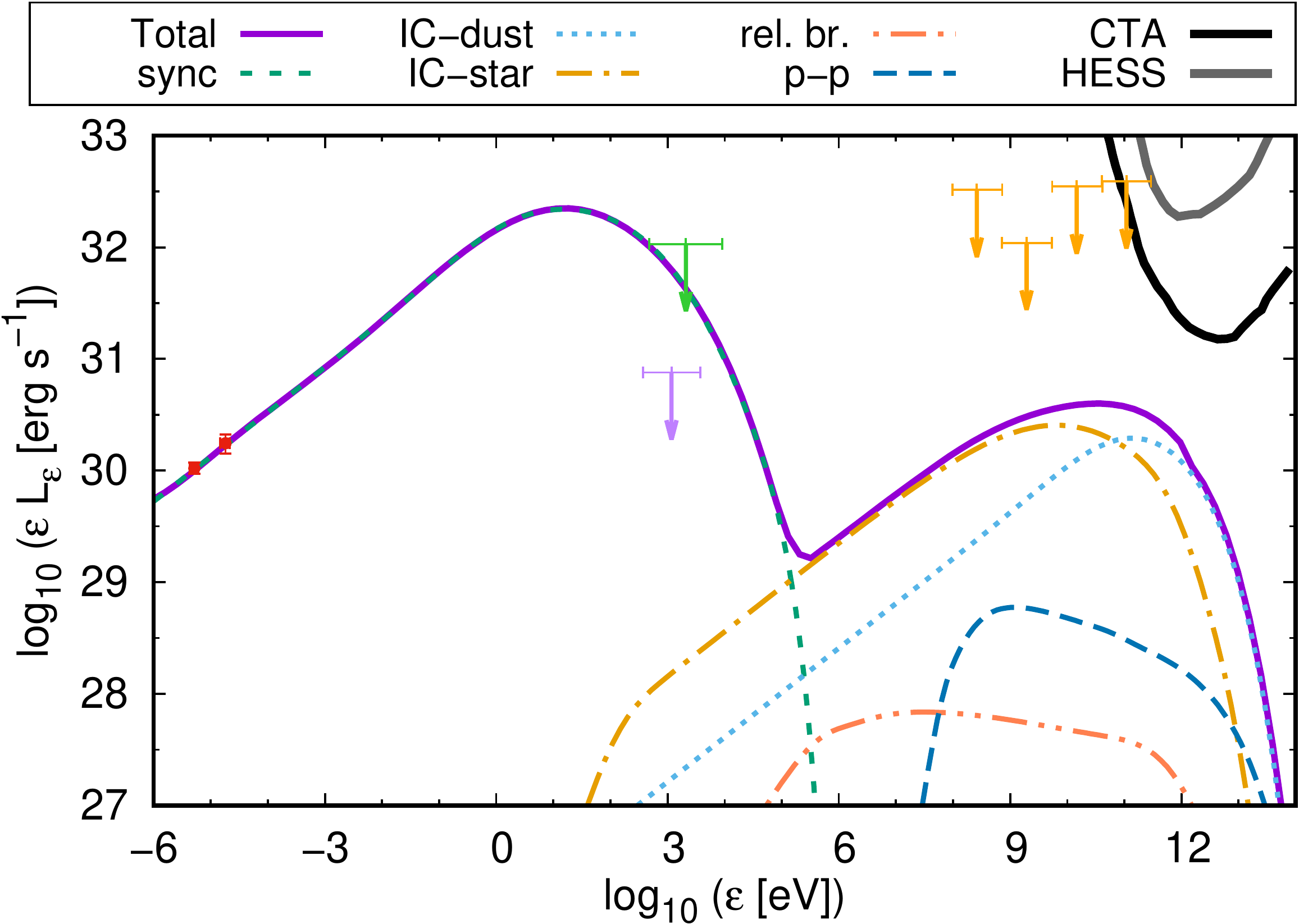}}   
    \caption[]{SED for a low (top) and a high magnetic field scenario (bottom). The red dots represent the VLA flux \citep{Benaglia2010}, the green arrow is the \textit{Suzaku} upper-limit (UL) in 0.3--10~keV \citep{Terada2012}, the purple arrow the \textit{XMM-Newton} UL in 0.4--4~keV \citep{Toala2016}, the orange arrows
    the \textit{Fermi} UL in 0.1--300~GeV \citep{Schulz2014}. The grey and black solid lines are the instrument sensitivities for 100-h HESS and 100-h CTA, respectively \citep{Funk2013}.} 
    \label{fig:SEDs}
  \end{figure}

The calculated broadband SEDs are shown in Fig.~\ref{fig:SEDs}, along with some instrument sensitivities and observations 
from different epochs. The integrated luminosities for each process are 
\mbox{$L_\mathrm{sy} \approx 8 \times 10^{32}$~erg~s$^{-1}$}, 
\mbox{$L_\mathrm{Br} \approx 2 \times 10^{30}$~erg~s$^{-1}$},
\mbox{$L_\mathrm{IC,dust} \approx 5 \times 10^{32}$~erg~s$^{-1}$}, 
\mbox{$L_\mathrm{IC,\star} \approx 10^{33}$~erg~s$^{-1}$}, and 
\mbox{$L_\mathrm{pp} \approx 4 \times 10^{29}$~erg~s$^{-1}$}.
In this case $n_\mathrm{ssw} \sim 0.04$~cm$^{-3}$ and $n_\mathrm{ISM} \sim 15$~cm$^{-3}$, so relativistic Bremmstrahlung and p-p emission could be enhanced more than three orders of magnitude if mixing is efficient. This 
would be in tension with the observational upper limits to the $\gamma$-ray flux,
therefore suggesting that efficient mixing is not likely and that the laminar approximation of the fluid is consistent.

We predict that, in the most favourable case, the system might be detectable at HE by the \textit{Fermi} satellite. At very HE a detection should wait until the Cherenkov Telescope Array (CTA) becomes operational.

\subsubsection{Scenario with equipartition magnetic field} \label{subsec:highB}

We apply the same model as in Sect.~\ref{subsec:lowB}, but with a different value of $\zeta_B$ in order to analyse a
scenario with an extremely high magnetic field. We explore the case of $\zeta_B= 1$, which corresponds to equipartition
between the magnetic field and the thermal pressure in the shocked SW (Eq.~\ref{eq:B}). Under such
conditions the magnetic field would be dynamically relevant (i.e. the magnetic pressure would be a significant fraction
of the total pressure in the post-shock region). Nonetheless, we do not alter our prescription of the flow properties, as
we adopt a phenomenological model for them only to get a rough approximation of the gas properties in the shocked SW.
Besides, our intentions are just to give a semi-qualitative description of this extreme scenario, not to make a precise
modelling of the emission, and to obtain a rough estimate of some relevant physical parameters. Fixing $\zeta_B= 1$
yields $B \sim 100$~$\mu$G in the shocked SW, and, if the magnetic field in the BS is solely due to adiabatic compression
of the stellar magnetic field lines, we get $B_\star < 850$~kG on the stellar surface. Such high values of
$B_\star$ are uncommon \citep{Neiner2015}, which suggests that in the energy equipartition scenario the magnetic field in
the BS is more likely generated or at least amplified \textit{in situ}. Setting $f_\mathrm{NT,e} \approx 0.004$ leads to
the spectral fit of the fluxes obtained by \citet{Benaglia2010} shown in Fig.~\ref{fig:SEDs}. There is a tension 
between the X-ray flux upper limits by \citet{Toala2016} and the synchrotron emission that extends to energies above
1~keV. This could be either evidence that the magnetic field in \object{BD$+43^{\circ}3654$} is not so extreme, or
that the particle acceleration efficiency is overestimated. The integrated luminosities for each process for this case are 
\mbox{$L_\mathrm{sy} \approx 1.7 \times 10^{33}$~erg~s$^{-1}$}, 
\mbox{$L_\mathrm{Br} \approx 4.4 \times 10^{28}$~erg~s$^{-1}$}, 
\mbox{$L_\mathrm{IC,dust} \approx 1.3 \times 10^{31}$~erg~s$^{-1}$}, 
\mbox{$L_\mathrm{IC,\star} \approx 2.8 \times 10^{31}$~erg~s$^{-1}$}, and 
\mbox{$L_\mathrm{pp} \approx 4.3 \times 10^{29}$~erg~s$^{-1}$}.
  
As the synchrotron cooling time is shorter than the IC cooling time, the bulk of the NT emission is 
produced in the form of low-energy ($< 1$~keV) photons, and much less luminosity goes into $\gamma$-ray 
production. The value of $L_\gamma$ 
obtained is $\sim 10^2$ times smaller than the one obtained in Sect.~\ref{subsec:lowB}. This result seems 
consistent as, roughly, $L_\mathrm{sy} \propto f_\mathrm{NT,e} \times \zeta_B$, and we are considering a value 
$\sim 10^2$ times larger for $\zeta_B$, and therefore $f_\mathrm{NT,e}$ must be $\sim 10^2$ smaller in order to fit the observations. This thus explains that $L_\mathrm{IC} \propto f_\mathrm{NT,e}$ is a factor $\sim 10^2$ smaller. In this 
scenario \object{BD$+43^{\circ}3654$} is not detectable with the current or forthcoming gamma-ray observatories.

\subsubsection{Resolved emission} \label{sec:maps}

The full information from the spatially resolved radio observations is only partially contained in the SEDs. 
Therefore, we compare the morphology predicted by our model with that observed by \citet{Benaglia2010} using VLA
observations at two frequencies,1.42~GHz and 4.86~GHz. We use a synthesised beam of $12\arcsec \times 12\arcsec$
in our synthetic maps so it matches with the synthesized beam of the VLA observations from \citet{Benaglia2010}, with
a corresponding resolution of $28\arcsec \times 28\arcsec$ (the relation between resolution and beam size is given in
Sect.~\ref{sec:synthetic_maps}).

\begin{figure*}
    \centering
    \hspace*{-1.5cm} \includegraphics[width=0.85\textwidth, angle=0]{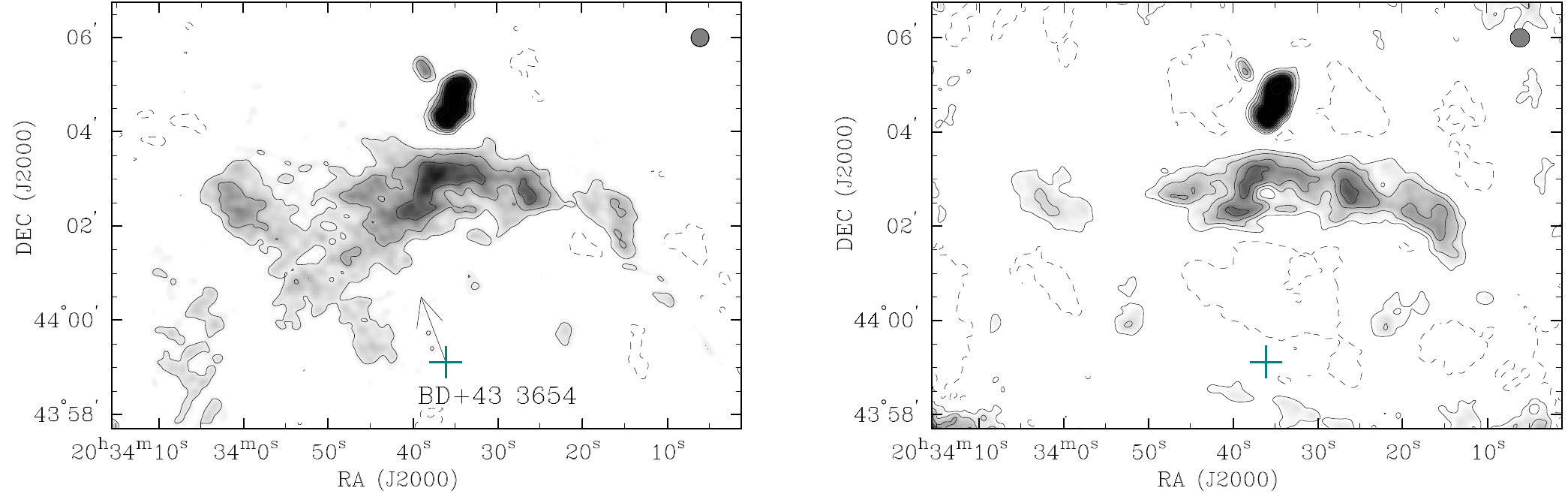}
    \includegraphics[width=0.28\textwidth, angle=270]{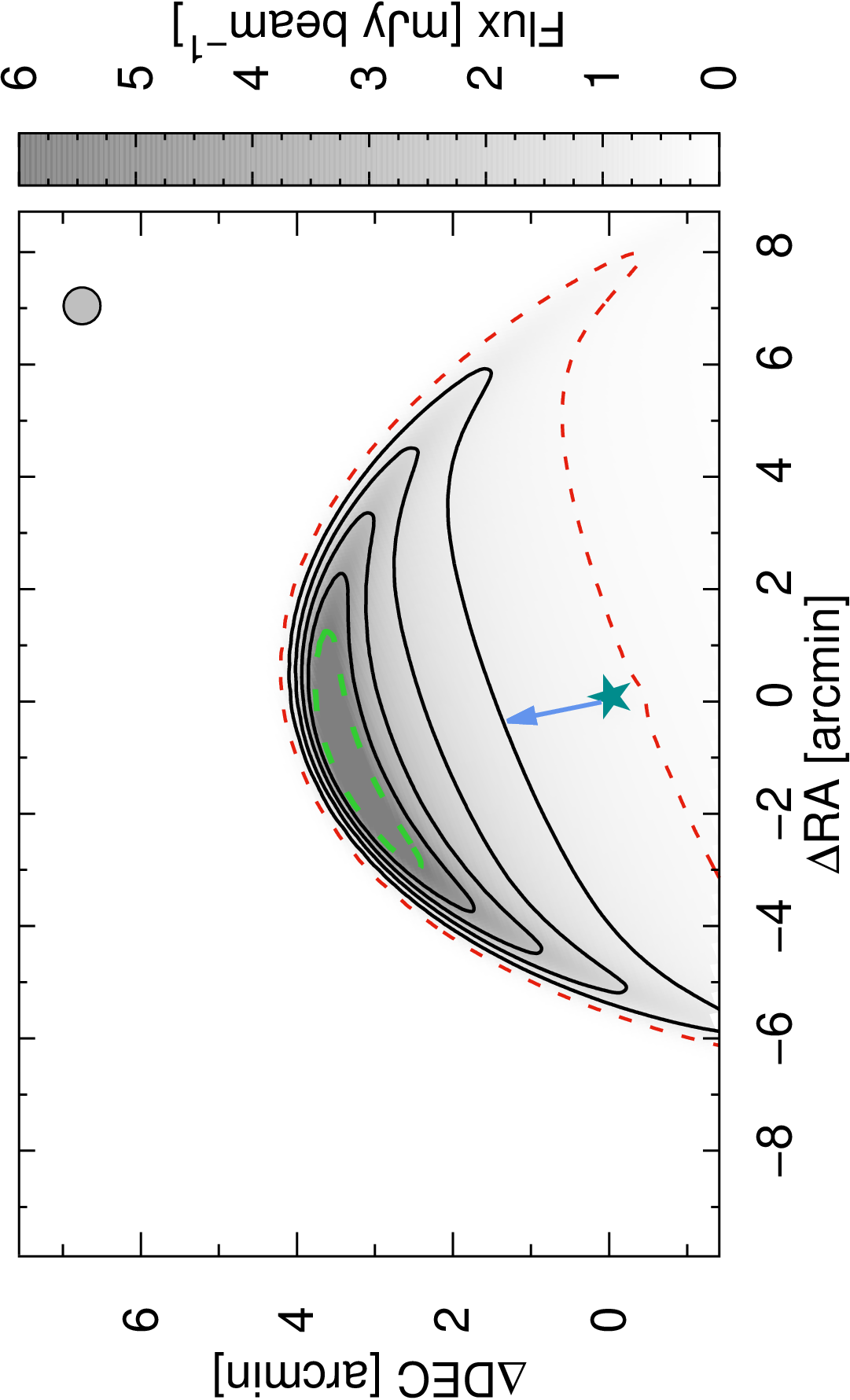}
    \includegraphics[width=0.28\textwidth, angle=270]{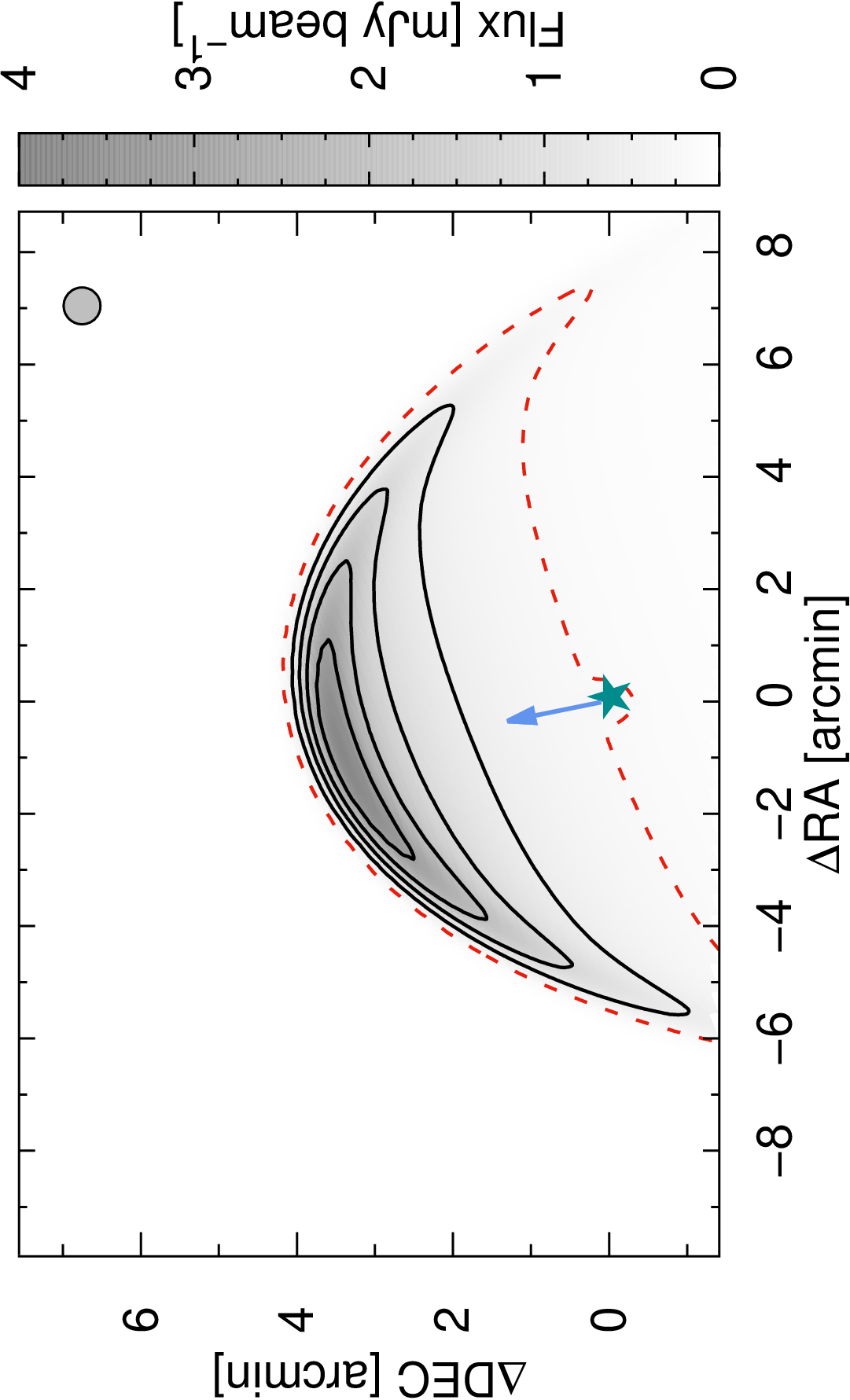}
    \caption[]{Comparison between the observed radio emission maps taken from \citet{Benaglia2010} (top) 
    and our synthetic maps (bottom). In the top right corner a synthesised beam of 
    $12\arcsec \times 12\arcsec$ is shown (i.e. $\sigma_x = \sigma_y = 12\arcsec$). Left and right panels
    correspond to an observing frequency of 1.42~GHz and 4.86~GHz, respectively. The rms levels of the
    observed maps are 0.3~mJy~beam~$^{-1}$ (left) and     
    0.2~mJy~beam$^{-1}$ (right). The black solid contours of the 1.42~GHz maps are at 0.9,1.8, 3.0, and    
    4.5~mJy~beam$^{-1}$, the red dotted contour is at 0.3~mJy~beam$^{-1}$ (the observed map rms), and 
    the green contour is at 6~mJy~beam$^{-1}$ (above the observed values). In the 4.86~GHz
    maps, the black contours are at 0.6, 1.2, 2.0, and 3.0~mJy~beam$^{-1}$, and the red dotted is at 0.2 mJy~ beam$^{-1}$ (the observed map rms). A projection factor $\cos{(DEC)}$ was used for the $x$-coordinates 
    in the synthetic map in order to relate the synthetic map units to sky positions.} 
  
    \label{fig:radio_maps}
\end{figure*}

As shown in Sect.~\ref{sec:synthetic_maps}, the morphology of the emission map depends on the inclination angle $i$.
Based on the observed morphology, we can argue that $45\degr < i < 90\degr$. Figure ~\ref{fig:radio_maps} shows that
there is a good agreement between the synthetic map and the observed map for $i \sim 75\degr$, both in morphology and
emission levels. However, in the modelled map the emission is a bit more extended than the observed one. This could be
attributed, for instance, to a magnetic field that drops faster with the apex distance than what 
our model assumptions predict (Sect.~\ref{subsec:hydrodynamics}). A more detailed analysis of the magnetic field
structure could be addressed by assuming frozen-in conditions for the magnetic field in each individual line of fluid 
of the shocked SW. Also, MHD simulations of colliding-wind binaries performed by \citet{Falceta2012} suggest that the magnetic pressure is not a constant fraction of the thermal pressure throughout the shocked plasma. However, the 
implementation of detailed MHD models for the shocked fluid is beyond the scope of this work.


\section{Conclusions}\label{sec:conclusions}

   We show that one-zone models can be reconciled with observations by properly accounting for the intensity of the IR
   dust photon field and motion of the plasma along the shocked region. The multi-zone model we developed improves the
   constraints on the different model parameters, namely the magnetic field intensity and the amount of energy deposited
   in NT particles. Our model reproduces fairly well the only radio observations for the object
   \object{BD$+43^{\circ}3654$}. However, the free parameters, namely the fraction of available energy that goes into
   accelerating NT electrons and the magnetic field intensity along the shocked SW, can only be
   constrained by current facilities (radio interferometers, X-ray and $\gamma$-ray satellites) with deep, 
   high-sensitivity observations. Comparison between the synthetic and observed radio maps allows us to constrain the
   star direction of motion with respect to the observer. Discrepancies in the morphology could account for deviations in
   the system parameters and/or model hypothesis, such as a highly non-uniform environment or that the magnetic field
   pressure does not remain a constant fraction of the thermal pressure. Estimating the magnetic field strength in the
   shocked region allows us to set upper limits for the magnetic field on the stellar surface, thereby inferring whether
   magnetic field amplification is taking place in the particle acceleration region. 

 The results presented in this work provide a good insight for future observational campaigns in the radio and 
 $\gamma$-ray range. In particular, we show that the most relevant parameters for the radiative output are the mass-loss rate and velocity of the wind of the star, rather than the density of the ISM or the stellar spatial velocity. The NT luminosity (especially the $\gamma$-ray luminosity) is strongly dependent on the mass-loss rate,
 whereas the detailed shape of the SED is defined by the magnetic field strength, the ambient density, and the ratio 
 $\dot{M}/v_\mathrm{w}$ according to Eq.~\ref{eq:ratio_radio_gamma}. Therefore, deep observations in soft X-rays
 ($0.3-10$~keV) could provide tighter constraints to the free parameters in our model, such as the injected particle
 energy distribution spectral index and the magnetic field strength, through comparison of the synchrotron and IC
 emission, constraining then the acceleration efficiency. We show that moderate values of the stellar surface magnetic
 field ($B_\star<100$~G) are sufficient to account for the synchrotron emission even if there is no amplification of the
 magnetic field besides adiabatic compression; however, if the magnetic field in the BS region is high 
 ($\gtrsim 100$~$\mu$G), then magnetic field amplification is likely to occur. Low $B$-values would imply significant gamma-ray emission, whereas high $B$-values render the predicted gamma-ray radiation undetectable with present or forthcoming instrumentation.
     

\begin{acknowledgements}
We thank the anonymous referee for the detailed and constructive comments. 
This work is supported by CONICET (PIP2014-0338) and ANPCyT (PICT-2017-2865). SdP acknowledge support by PIP 0102 (CONICET). V.B-R. acknowledges support from MICINN by the MDM-2014-0369 of ICCUB (Unidad de Excelencia 'Mar\'ia de Maeztu'). V.B-R. and G.E.R. acknowledge support by the Spanish Ministerio de Econom\'{\i}a y Competitividad (MINECO) under grant AYA2016-76012-C3-1-P. 
The acknowledgment extends to the whole GARRA group.
\end{acknowledgements}


\bibliographystyle{aa} 
\bibliography{biblio} 


\end{document}